\documentclass[10pt,journal,compsoc]{IEEEtran}
\ifCLASSOPTIONcompsoc

\ifCLASSINFOpdf
\usepackage[pdftex]{graphicx}
\graphicspath{{figures/}{pictures/}{images/}{./}} 
\DeclareGraphicsExtensions{.pdf,.png,.jpg,.jpeg}
\else
\fi
\usepackage{caption}

\graphicspath{{figures/}{pictures/}{icon/}{./}} 
\usepackage{microtype} 
\PassOptionsToPackage{warn}{textcomp}  
\usepackage{textcomp}  
\usepackage{mathptmx}  
\usepackage{times} 
\usepackage{tabu}  
\usepackage{booktabs}  
\usepackage{lscape}

\usepackage[T1]{fontenc}
\usepackage{dfadobe}  
\usepackage{picinpar}
\usepackage{wrapfig}
\usepackage{pdfpages}
\usepackage{ulem}
\usepackage{amsmath}
\usepackage{amssymb}

\usepackage{amssymb}
\usepackage{tabularx}
\usepackage[figuresleft]{rotating}
\usepackage{graphicx}
\usepackage{grffile}
\usepackage{subfigure}
\usepackage{multirow}
\usepackage{colortbl}
\usepackage{enumitem}
\usepackage{enumerate}
\usepackage{threeparttable}
\usepackage{makecell}
\usepackage{stfloats}
\usepackage{ragged2e}
\usepackage{listings}
\definecolor{bluecrayola}{rgb}{0.12,0.46,1.0}

 \newcommand{\review}[1]{\textcolor{brown}{}}

\title{Graph Exploration with Embedding-Guided Layouts}

\begin{document}

	\author{
	Leixian~Shen*,
	Zhiwei~Tai*,
	Enya~Shen,
	and~Jianmin~Wang
	
	\IEEEcompsocitemizethanks{
		\IEEEcompsocthanksitem 
		All authors are from Tsinghua University, Beijing, China.
		E-mail: 
		$\left\{slx20@mails., tzw20@mails., shenenya@, jimwang@\right\}$tsinghua.edu.cn.
		\IEEEcompsocthanksitem 
		* Both authors contributed equally to the paper.
	}
	\thanks{Manuscript received XX XX, 2023; revised XX XX, 2023.}}

\markboth{\MakeUppercase{LATEX},~Vol.~XX, No.~X, XX~2023}%
{Shell \MakeLowercase{\textit{et al.}}: Bare Demo of IEEEtran.cls for Computer Society Journals}

\IEEEtitleabstractindextext{
	\begin{abstract}
		\raggedright
		Node-link diagrams are widely used to visualize graphs. Most graph layout algorithms only use graph topology for aesthetic goals (e.g., minimize node occlusions and edge crossings) or use node attributes for exploration goals (e.g., preserve visible communities). Existing hybrid methods that bind the two perspectives still suffer from various generation restrictions (e.g., limited input types and required manual adjustments and prior knowledge of graphs) and the imbalance between aesthetic and exploration goals. In this paper, we propose a flexible embedding-based graph exploration pipeline to enjoy the best of both graph topology and node attributes. First, we leverage embedding algorithms for attributed graphs to encode the two perspectives into latent space. Then, we present an embedding-driven graph layout algorithm, GEGraph, which can achieve aesthetic layouts with better community preservation to support an easy interpretation of the graph structure. Next, graph explorations are extended based on the generated graph layout and insights extracted from the embedding vectors. Illustrated with examples, we build a layout-preserving aggregation method with Focus+Context interaction and a related nodes searching approach with multiple proximity strategies. Finally, we conduct quantitative and qualitative evaluations, a user study, and two case studies to validate our approach.

	\end{abstract}
	
	}
\maketitle
\IEEEdisplaynontitleabstractindextext
\IEEEpeerreviewmaketitle

\maketitle
\section{Introduction}
Graphs are widely used to encode data with topology structures (e.g., biological networks~\cite{Bharadwaj2022,Gehlenborg2010}, social networks~\cite{Wu2016,Bezerianos2010}, and deep learning dataflows~\cite{wongsuphasawat2017visualizing}).
Compared with numerical assessment, visualizing graphs as node-link diagrams can depict the overall graph structure for more intuitive and efficient analysis. 

Layout is a fundamental task for node-link diagram exploration. 
Over the years, various promising layout methods have been designed to visualize graphs, such as force-directed approaches~\cite{fruchterman1991graph,suh2019persistent}, dimensionality reduction-based algorithms~\cite{Kruiger2017,Brandes2007,Zhu2021}, deep learning-based methods~\cite{wang2019deepdrawing,2019arXiv190412225K}, and multi-level approaches~\cite{Zhu2021,Hachul2005}.
However, most layout methods are either topology-driven or attribute-driven, which focus too much on one side for a certain goal.
Graph topology contains connection information between nodes and is important for aesthetic presentation (e.g., with few node occlusions and edge crossings)~\cite{fruchterman1991graph,haleem2019evaluating,dunne2015readability}.
Node attributes incorporate various additional information about the node and are usually used for graph clustering, which partitions nodes into disjoint communities, and nodes of the same community share some commonalities~\cite{Itoh2015}. 
A pure attribute-driven layout can be overly compact as it does not consider the topology feature. Ignoring the node attributes may miss the important community information for analytic goal.
In comparison, integrating the two crucial elements for graph layout has received relatively little attention in current research~\cite{gibson2013survey,Fischer2021,Chen2019g,nobre2019state,Kerren2014}.




According to our survey, there have been several prior efforts that attempt to integrate the two perspectives for graph visualization. 
However, these hybrid methods have two major issues in terms of layout quality and generation restriction.
For the layout quality, most middle-ground approaches combine the two elements in a hierarchical form~\cite{Itoh2015,Shi2014,Archambault2008a}. For example, Itoh \textit{et al.}\cite{Itoh2015} and OnionGraph~\cite{Shi2014} first use node attributes to calculate clusters and then adopt a layout algorithm to position the clusters. The target of community preservation prevails in these approaches, and aesthetic goals play a relatively weak role. Unlike the loosely-coupled binding style, GraphTSNE~\cite{Leow2019} leverages Graph Convolutional Network (GCN) with a modified \textit{t}-SNE loss to encode graph connectivity and node attributes together. However, the produced layouts can not reveal visible communities, which will be further discussed in the evaluation.
For the generation restriction, some approaches depend on user interaction to adjust the layout~\cite{Spritzer2008,Spritzer2012,Jusufi2013}, which is usually a time-consuming process.
For instance, MagnetViz~\cite{Spritzer2012} allows users to manipulate virtual magnets, which represent a particular attribute and can attract nodes that meet a set of associated criteria. However, only a few attributes can be applied in one manipulation, and it requires users' prior knowledge of the graph structure.
In addition, many methods also have specific input restrictions. For example, GraphTSNE~\cite{Leow2019} can only accept graphs with multiple attributes as input but fail to handle nodes with a single or without attribute.   MVN-Reduce~\cite{Martins2017} and JauntyNets~\cite{Jusufi2013} only handle quantitative attributes and are only for dimensionality reduction-based and force-based layout algorithms, respectively.
In general, the above attempts still have shortcomings about layout quality  (e.g., imbalance between aesthetic and exploration goals) and generation restriction (e.g., limited input types and required manual adjustments and prior knowledge of graphs). 
So our target is to integrate graph topology and attributes in a flexible and organic manner to avoid the restrictions and better balance aesthetic and exploration goals.

Recently, graph embedding has proven effective in obtaining high-level feature vectors of graphs~\cite{liao2018attributed,gao2018deep,zhu2018deep}. An essential category is based on random walks (e.g., node2vec~\cite{grover2016node2vec} and DeepWalk~\cite{ozzi2014deepwalk}), which represents the graph as a set of sampled random walking paths.
These methods usually require less computational cost and are proven flexible and effective.
This inspires us to leverage graph embedding to bind topology and node attributes for layout and graph exploration.

\begin{figure*}[t]
	\centering 
	\includegraphics[width=0.9\textwidth]{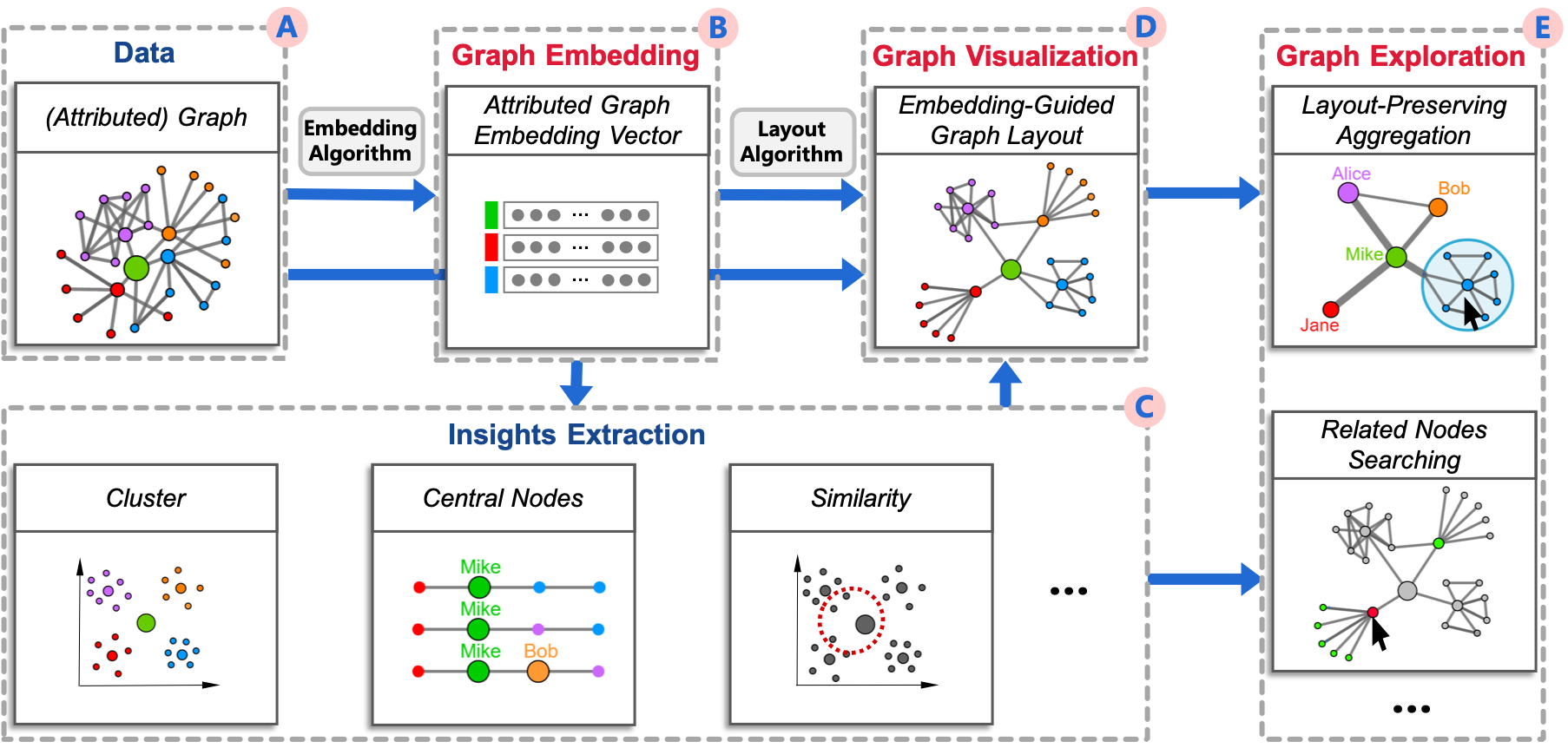}
	\caption{
		Embedding-based graph exploration pipeline. The graph embedding algorithm encodes attributed graphs into low-dimensional vectors, from which can extract rich data insights (e.g., cluster, central nodes, and similarity. 
		The embedding-guided layout method integrates similarities and connections between nodes to produce aesthetic and community-aware layouts.
		The generated layout, coupled with the extracted insights, enables various graph exploration applications (e.g., node aggregation and related nodes searching).
	}
	\label{fig:pipeline}
\end{figure*}

In this work, we introduce a flexible pipeline (Fig.~\ref{fig:pipeline}) for undirected graph exploration that leverages graph embedding to extract high-level features of attributed graphs, power better layout results, and guide graph exploration applications. 
Layout plays a vital role in the pipeline to visualize the underlying data insights for further exploration. In the pipeline, we include an embedding-driven graph layout algorithm, GEGraph (GE is short for Graph Embedding). GEGraph combines the similarity information between nodes in latent space and the connection information of graph structure to produce visually appealing (e.g., with few node occlusions) and community-aware graph layouts, balancing aesthetic and exploration goals. In addition, GEGraph is self-automatic and can adapt to graphs with different attribute types.
Based on the high-quality layout, we integrate the data insights extracted from the embedding vectors to design two exploration examples: layout-preserving graph aggregation and multi-strategy related nodes searching.
In general, our contributions are summarized as follows: 
\begin{itemize}[leftmargin=*]
\item 
We propose a flexible embedding-based graph exploration pipeline that extracts high-level features and data insights of attributed graphs to power layout and exploration.
\item 
We design an embedding-driven graph layout algorithm, which can bind graph topology and node attributes to produce aesthetic and community-aware layouts. 

\item 
We evaluate the effectiveness of our approach quantitatively and qualitatively by comparing it with popular layout methods, as well as a user study and two case studies with real-world data. 
\end{itemize}

\section{Related work}
Our work draws upon existing advances in three perspectives, i.e., graph layout, graph embedding, and graph exploration. 

\subsection{Graph Layout}
Node-link diagram layout algorithms can be briefly divided into topology-driven, attribute-driven, and hybrid approaches~\cite{nobre2019state}.

\textbf{Topology-driven}.
Topology-driven layout algorithms draw graphs based solely on graph topology. 
One of the classic families is force-directed, which generates a force-balanced graph layout by modeling nodes and edges as physical objects.
One representative example is the Fruchterman-Reingold (F-R) algorithm~\cite{fruchterman1991graph}, which defines a simplified charge-spring model and performs well with simplicity and scalability.
Abundant algorithms are later proposed by encoding extra information into the physical model~\cite{kk1989graph,graphSGD2018,noack2003energy,Bannister2013,suh2019persistent,Xue2022}. 
For example, most recently, PH~\cite{suh2019persistent} leverages the persistent homology features of graphs for interactive layout manipulation with a novel bar-list design.
Another type is dimensionality reduction-based methods, which leverage dimensionality reduction algorithms to project data onto a 2D space. For instance, PivotMDS~\cite{Brandes2007} is a sampling-based approximation to classical multidimensional scaling (MDS) by assigning a subset of nodes as pivots. HDE~\cite{Harel2002} is similar to PivotMDS but uses principal component analysis (PCA) for dimensionality reduction. Recently, \textit{ts}NET leverages \textit{t-}distributed Stochastic Neighbor Embedding (\textit{t-}SNE)~\cite{Kruiger2017} with a modified cost function to encode the graph as a 2D distance matrix. 
In recent years, advances in deep learning are introduced to power automatic graph layout, which performs well on learning the graph topology and drawing style~\cite{wang2019deepdrawing,2019arXiv190412225K,Kwon2018}. However, they are usually limited by the graph size and are not scalable to solve universal problems.
In addition, aiming at reducing computation time, multi-level methods decompose large graphs into coarser graphs. 
They first lay out the coarsest graphs with a force-directed ~\cite{Walshaw2001,Hachul2005,Gajer2001} or dimension reduction-based method~\cite{Cohen1997,Zhu2021} and then use the vertex position as the initialization for the next finer graph. 
Most recently, DRGraph~\cite{Zhu2021} enhances the non-linear dimensionality reduction scheme using a sparse distance matrix, the negative sampling technique, and a multi-level layout scheme. The scheme comprises coarsening, coarsest graph layout, and refinement.
In general, although topology-driven algorithms are effective, they only consider the graph topology  but ignore its important node attributes.

\textbf{Attribute-Driven}.
Attribute-driven layouts adopt positions of nodes to encode attribute information~\cite{nobre2019state}. 
The most common strategy is to alter the visual appearance (e.g., color, size, and shape) of the graph elements through on-node or on-edge encoding. For instance, Neuweger \textit{et al.}~\cite{Neuweger2009} make use of on-node encoding by means of color, where metabolic pathways are overlaid with multiple attributes (metabolite concentrations). There are also a set of biological systems that adopt embedded charts (e.g., bar charts, box plots, etc.) as nodes to visualize multivariate graphs~\cite{Gehlenborg2010}.
Faceting is also a common case in attribute-driven layout, which groups nodes by a categorical attribute and places nodes in each group with other methods~\cite{Pretorius2008Visual, 10.1109/TVCG.2013.223,10.5555/2384008.2384030, 10.1109/TVCG.2008.117, Partl2014ConTour}.
For instance, Shneiderman \textit{et al.}~\cite{SS2006BA} compute graph layouts by semantic substrates, which are non-overlapping regions, and the node placement is based on the attributes. 
Another case is attribute-driven positioning, which sets the nodes' position by parts of attributes (often numerical)~\cite{edgemaps2011}. For example, 
GraphDice~\cite{Bezerianos2010} proposes the scatterplot matrix design to manipulate attributes in social networks. Doerk \textit{et al.}~\cite{edgemaps2011} set the node positions in space by attribute-based dimension reduction. Additionally, graphTPP~\cite{Gibson2017,Gibson2016} emphasizes using attributes and clustering for graph layout, which can achieve a clear clustered structure.
Attribute-driven positioning is also commonly used in spatial networks~\cite{PMID:26356945,PMID:19834170} and graphs with 1D attribute (e.g., time~\cite{edgemaps2011} and genomic coordinates~\cite{Meyer2009MizBee, PMID:19541911}).
Generally, attribute-driven methods mostly make superficial use of attribute information and neglect the vital topological structure.

\textbf{Hybrid.}
Hybrid approaches attempt to use comprehensive information to aid graph layout. For example, MagnetViz~\cite{Spritzer2012} arranges graph nodes with a force-directed algorithm and provides users with virtual magnets, which acts as a specific attribute and can attract nearby nodes that meet certain associated criteria.
JauntyNets~\cite{Jusufi2013} places attribute nodes on a circle around the graph. Topological nodes with attribute values above a specific threshold are linked with the corresponding attribute nodes via node-to-attribute edges to integrate the attribute information into the force-based layout.
However, the quality of layouts depends on users' modification of parameters, and the two methods require users' prior knowledge of the graph.
GraphTSNE~\cite{Leow2019} trains a Graph Convolutional Network (GCN) on a modified \textit{t}-SNE loss. It accounts for both graph connectivity and node attributes but suffers from high training costs and input restrain (must have multiple attributes). 
MVN-Reduce~\cite{Martins2017} combines the two aspects into a unified model to act as the input of dimensionality reduction-based approaches. However, it only considers quantitative attributes. Similarly, JauntyNets can only handle numerical attributes and only apply to force-based approaches.
In addition to general layouts, hybrid approaches are also used to generate other structural information. In detail, Itoh \textit{et al.}~\cite{Itoh2015} combine structural neighborhood and attribute similarity to generate key-node-separated graph clusters and leverage the stress minimization layout algorithm to calculate the positions of vertices.
OnionGraph~\cite{Shi2014} enables node aggregation based on either attribute, topology, or their hierarchical combination.
Wu and Takatsuka~\cite{Wu2006,Wu2008} leverage spherical Self-Organizing Map (SOM) to group nodes with similar attributes to adjacent areas on the circular layout. 
Although these approaches are a promising start to combine the topology and attribute information, they still have drawbacks in terms of layout quality and generation restriction. 
In this paper, we attempt to propose a flexible graph exploration pipeline based on graph embedding to avoid the generation restrictions and produce more attractive and community-aware visualizations.

\subsection{Graph Embedding}

Existing embedding methods can be divided into structure-first and attribute-first~\cite{cai2018comprehensive}. 

\textbf{Structure-first}. DeepWalk~\cite{ozzi2014deepwalk} first leverages random walk to convert a graph into paths and then adopts the language model to generate node embedding.
LINE~\cite{tang2015line} learns node representations by modeling the first-order and second-order proximity.
Struc2vec~\cite{ribeiro2017struc2vec} encodes the role similarity of node structure into a multi-layer network, where the weight of edges in each layer is determined by the structural role difference of the corresponding scale.
Role2vec~\cite{ahmed2018learning} captures the behavioral roles of nodes with structurally similar neighborhoods (e.g., connectivity and subgraph patterns).
Node2vec~\cite{grover2016node2vec} presents a new random walk strategy to interpolate between breadth-first sampling and depth-first sampling. Though these embedding methods can effectively convert graphs into high-level vectors, they only consider the graph topology.

\textbf{Attribute-first}.
TADW~\cite{yang2015network} extends DeepWalk to incorporate a node-context matrix. HSCA~\cite{zhang2016homophily} integrates homophily, topology structure, and node content information to ensure effective network representation learning. However, the two matrix factorization-based methods are time-and-space-consuming when dealing with large feature matrices. 
SNE~\cite{liao2018attributed} includes two deep neural graph models which are used to process structure and attribute information in the embedding layer.
DANE~\cite{gao2018deep} designs two deep neural graph models to separate the topology and node attributes and then applies joint distribution to optimize the result.
DVNE~\cite{zhu2018deep} learns a Gaussian distribution in the Wasserstein space as the latent node representation.
BANE~\cite{yang2018binarized} defines a Weisfeiler-Lehman proximity matrix to capture data dependence between node links and attributes.
MetaGraph2Vec~\cite{Zhang2018c} and Metapath2vec~\cite{Dong2017} perform scalable representation learning in heterogeneous information networks through meta-path-guided random walks.
The above methods can capture richer graph feathers but require a long training time and lacks good scalability.

Considering the diverse graph visualization and exploration scenarios that depend on graph topology and node attributes at different levels, we choose a flexible structure-first embedding approach, node2vec, and extend it to integrate node attributes in the random work process.

\subsection{Graph Exploration}
There are various tasks associated with graphs (e.g., correlate, identify, compare, and categorize)~\cite{lee2006,Kerren2014}. With the embedding vectors, we can extract various intent-oriented insights and design interactive applications for data exploration.

For example, in high level, aggregation is a practical approach to reduce the complexity of graph layout by creating an overview of the graph with cluster information~\cite{nobre2019state}. 
PivotGraph~\cite{wattenberg2006visual} aggregates nodes by the attributes, and the size of aggregated nodes reflects the number of nodes with such attributes.
Elzen and Wijk~\cite{van2014multivariate} design an approach to provide an overall view for large geographical graphs.
Tensorflow~\cite{wongsuphasawat2017visualizing} aggregates the mathematical operations or functions to a module to help increase the interpretability of users' models.
ASK-GraphView~\cite{VanHam2006} enables clustering and interactive navigation of large-scale graphs.
To interactively explore the aggregated nodes, Herman et al.~\cite{herman2000graph} introduce Focus+Context, which aims to show the overall and detailed view in the same area to reveal the graph structure. In addition, related nodes searching is a common task in graph exploration with the similarity information.
Chen \textit{et al.}~\cite{PMID:30136986} propose a structure-based suggestive exploration approach by encoding nodes with vectorized representations.
It can identify similar structures in a large network, by which users can interact with multiple similar structures. 
However, it only encodes the graph topology information. We propose an embedding-based approach with three searching strategies.

Our pipeline supports highly customizable graph exploration applications based on the embedding vectors and graph layout. We illustrate the scalability of the pipeline with two examples: graph aggregation and related nodes searching. 
Furthermore, Pretorius \textit{et al.}~\cite{Kerren2014} present a framework of tasks for multivariate networks. We hope that our pipeline and designed examples can inspire more interesting graph exploration applications for these analytic tasks, even with different interaction modalities~\cite{Srinivasan2018,Shen2021a,Buschel2019}.


\section{Pipeline}\label{sec:pipeline}
In this paper, we contribute an embedding-based pipeline for graph exploration that considers both graph topology and node attributes, as shown in Fig.~\ref{fig:pipeline}. 
The pipeline is flexible, working as follows: we first leverage graph embedding algorithms to convert attributed graphs (A) into low-dimensional vectors (B), which support more complex data transformation types. Then a set of valuable data insights (C) can be extracted from the embedding results such as community clusters, central nodes that are representatives for communities, similarity between nodes, etc. The insights are essential for the graph layout and exploration applications. Next, we design an embedding-driven graph layout algorithm, GEGraph, that integrates the similarity matrix generated from the embedding vectors with the adjacency matrix of the graph topology to enhance existing graph layout method to achieve aesthetic and community-aware graph layouts (D). With the generated layout and extracted insights at hand, we can develop various interesting applications (E) for interactive graph exploration such as node aggregation and related nodes searching. 

In our instantiation, for the graph embedding algorithm, we extend node2vec~\cite{grover2016node2vec}, a widely used topology-based embedding method, to node2vec-a (a is short for attribute), by integrating node attributes as virtual nodes into the random walk process. The design logic links the two elements in a flexible manner, which will be discussed in Section \ref{sec:embedding}. For the layout algorithm, we select the F-R algorithm~\cite{fruchterman1991graph} as a basis for the prototype of GEGraph due to its wide applications, ease of implementation and integration, which will be discussed in Section \ref{sec:layout}. On top of the generated layout, to demonstrate the usability of the exploration pipeline, we design a layout-preserving node aggregation application (with central nodes that are representatives for communities) and a multi-strategy related nodes searching application (with node similarity), which will be discussed in Section \ref{sec:exploration}.

The modules are loosely coupled, and the pipeline allows for a high degree of customisability. 
On the one hand, users can replace the embedding and layout algorithm with other advanced or proprietary approaches to satisfy their certain goals; on the other hand, users can apply various data mining methods to extract insights from the embedding-based vectors and design their interactive graph exploration applications by the intersection of multiple areas (e.g., visualization, human-computer interaction, and data science).
We also expect more interesting applications with different graph insights to be designed in the future.

\section{Attributed Graph Embedding}\label{sec:embedding}
In this section, we leverage attributed graph embedding to integrate graph topology and node attributes in a flexible manner and transform graph features into low-dimensional vectors for graph visualization and exploration.

\subsection{Problem Definition}
An attributed graph is formally denoted as $G=(V,E,\Lambda)$, where $V$ and $E$ are sets of nodes and edges, respectively. 
$\Lambda=\{\alpha_1,...,\alpha_m\}$ is the set of attributes associated with the nodes in $V$.
Each node $v_i$ in $V$ corresponds to a vector $[\alpha_1(v_i),...,\alpha_m(v_i)]$, where $\alpha_j(v_i)$ is the attribute value of $v_i$ on attribute $\alpha_j$.
The goal of embedding is to generate a mapping function from the bipartite graph to the feature representation $f:G\rightarrow \mathbb{R}^d$, where $d$ is the dimension number of feature vectors.

In graph visualizations, edges between nodes reflect the structural connections. Likewise, if two nodes have the same values on specific attributes, it is intuitive that these nodes have a semantic connection, which is an important relationship in graph visualization. So we believe that attributes can work as another kind of ``edge" in graphs. However, as the dependency on topology and attributes varies in diverse graph visualization and exploration scenarios, we confront a challenge in applying the embedding algorithm for graph visualization: how to flexibly extract graph features in different perspectives, such as local structure, global structure, and node attributes.

\begin{figure}[t]
	\centering 
	\includegraphics[width=0.36\textwidth]{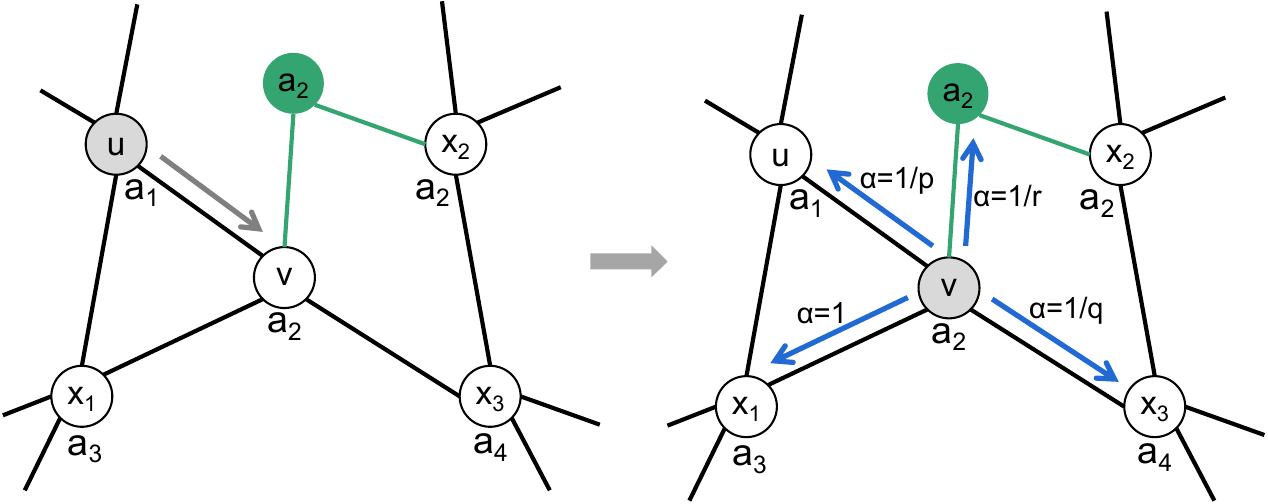}
	\caption{Illustration of modeling graphs with node attributes. Inspired by node2vec~\cite{grover2016node2vec},
		by setting $p$, $q$, and $r$, random walking paths based on local structure, global structure, and node attribute can be generated, revealing different proximities.
	}
	\label{fig:random_walk}
	\vspace{-0.3cm}
\end{figure}

\subsection{Graph Modeling}
Inspired by node2vec~\cite{grover2016node2vec}, a widely used graph embedding method, which introduces two notions of a node’s neighborhood for a flexible biased random walk procedure, we attempt to extend node2vec to node2vec-a by integrating attributes into the random walk process and providing flexibility in the transition between attribute-based and structure-based features for graph visualization.

To this end, we model the graph by extending the definition of node and edge.
As shown in Fig.~\ref{fig:random_walk}, the attributed graph is extended by taking attributes $\{\alpha_1,..., \alpha_m\}$ as the virtual nodes $\{v_{\alpha 1},...,v_{\alpha m}\}$: if node $v_i$ has attribute $\alpha_j$, there will generate a virtual edge between $v_i$ and $v_{\alpha j}$. The extended graph is denoted as $G'= (V', E', \Lambda')$, where $V'$ is the union set of real nodes and virtual attribute nodes, and $E'$ is the union set of real edges and virtual edges.
$\Lambda'$ is the same as $\Lambda$. 
Take attribute $a_2$ in Fig.~\ref{fig:random_walk} as an example, the attribute $a_2$ shared by $v$ and $x_2$ is added into the graph as a virtual node $a_2$, and edges $e_{v,a_2}$ and $e_{x_2,a_2}$ are established. After the first step walking from $u$ to $v$, subsequent steps are computed according to walking parameters.

In addition, the node attributes can be mainly divided into nominal and quantitative types. Nominal attributes can be directly converted into virtual nodes in the model. For quantitative attributes, we leverage the Chi Merge algorithm~\cite{Lehtinen2012} to discretize them into bins and adopt the mean value to represent attributes in the bin. Then they can be processed as nominal ones.

\subsection{Random Walk Strategy}\label{sec:strategy}
Given a source node $u$ and a fixed random walk length $l$, the $i$th node in the path $c_i$, which starts with $c_0 = u$, is generated by the following distribution:

\begin{equation}
	P(c_i=x|c_{i-1}=v) = \begin{cases}
		\frac{\pi_{vx}}{Z}, & if (v, x) \in E'\\
		0, & \text{otherwise}	
	\end{cases}
\end{equation}

where $\pi_{vx}$ is the unnormalized transition probability between nodes $v$ and $x$, and $Z$ is the normalizing constant. 
Since $E'$ includes extended attribute edges, the random walk will take node attributes into account.
A significant problem here is how to define the probability of random walk for virtual attribute nodes.

Let $V_\Lambda\subset V'$ denote the set of virtual attribute nodes in $G'$, and $\alpha$ denote the search bias of the next step from the source node $u$ to the target node $x$. Formally, inspired by node2vec~\cite{grover2016node2vec}, the transition probability $\pi_{vx}$ is designed as:

\begin{equation}
	\pi_{vx} = \begin{cases}
		\frac{1}{r}, & if ~v~ or~ x \in V_{\Lambda}\\
		\alpha(v, x), &\text{otherwise}	
	\end{cases}
\end{equation}

\begin{equation}
	\alpha(v, x) = \begin{cases}
		\frac{1}{p}, & if~ d_{ux}=0\\
		1, & if~ d_{ux}=1\\
		\frac{1}{q}, & if~ d_{ux}=2	
	\end{cases}
\end{equation}

where $p$ controls the likelihood of walking to local structure nodes that are interconnected and belong to the same community under the homophily hypothesis~\cite{Yang2014f}, $q$ controls the likelihood of walking to global structure nodes that play similar structural roles in different communities under the structural equivalence hypothesis~\cite{Henderson2012}, and $r$ decides the bias between topology and node attributes, which extends the definition of graph in node2vec~\cite{grover2016node2vec}. $d_{ux}$ is the shortest distance between the source node $u$ and node $x$.

Generally, we further divide the proximity in the random walking process between nodes into three categories:
\begin{itemize}[leftmargin=*]
	\item \textbf{Local structural proximity:} 
	If $p<min(q, r, 1)$, the probability of returning from the source node to the previous step increases. Thus, the random walking process is closer to local structure-based, which encourages moderate exploration.
	
	\item \textbf{Global structural proximity:} 
	If $q<min(p, r, 1)$, the probability of walking from the source node to other distant real nodes increases. Thus, the random walking process is closer to global structure-based, which encourages outward exploration.
	
	\item \textbf{Attribute proximity:} 
	If $r<min(p, q, 1)$, the probability of walking from the source node to virtual attribute nodes increases. Thus, the random walking process is closer to attribute-based, which encourages property-preserving exploration.

\end{itemize}

By adjusting walking parameters $p$, $q$, and $r$, the topology and attribute information can be encoded into low-dimensional vectors with different priorities in the embedding process, revealing different proximities for diverse visualization and exploration scenarios. Finally, we adopt the Skip-gram model to learn the latent representation of the graph.

\section{Embedding-driven Graph Layout}
\label{sec:layout}

After embedding, both the topology and attribute information are mapped into feature vectors.
This section will discuss how the graph layout approach, GEGraph, evolves from the F-R algorithm~\cite{fruchterman1991graph} to enhance graph layout with embedding results.

\subsection{F-R Algorithm}\label{sec:fr}
F-R~\cite{fruchterman1991graph} is a typical instance of the force-directed layout algorithm.
The design of attractive and repulsive force is the core of force-directed layout algorithms.
In the F-R algorithm, attractive force $f_a(d)$ is calculated as:
\begin{equation}
	f_a(d) = \frac{d^2}{k}
\end{equation}
where $d$ is the current distance of two nodes, and $k$ is a constant to control the minimal gaps between nodes.
Repulsive force $f_r(d)$ is calculated as:
\begin{equation}
	f_r(d) = -\frac{k^2}{d}
\end{equation}
Applying the F-R algorithm is an iterative process, and the resultant force applied on each node makes an offset of position in each iteration. It repeats until the nodes achieve convergence.
The F-R algorithm can generate visually appealing layouts and is easy to implement and integrate with our method. So we choose F-R as the foundation for the prototype.
Fig. \ref{fig:basic_layout} (a) shows the output of the F-R algorithm with the Les Mis{\'e}rables~\cite{suh2019persistent} dataset, which is the baseline of our layout algorithm design.

\subsection{Dimension Reduction of Feature Vectors}

The graph layout process can be viewed as computing a 2D dimensional position vector for each node.
So we first try to directly leverage dimension reduction technologies to map the embedding results into a 2D space.
Inspired by recent work~\cite{Zhou2021}, we apply $t$-SNE~\cite{vandermaaten08a} to the high-dimensional vectors.
As Fig. \ref{fig:basic_layout} (b) shows, the resulting layout can reveal proper node distribution information, where node communities can be distinguished intuitively.
However, edges in the graph are mostly crossed with each other inside communities, making the layout less aesthetic than Fig. \ref{fig:basic_layout} (a). Due to the fact that whether nodes are linked or not is ignored in the layout process, the edge crossing problem in Fig. \ref{fig:basic_layout} (b) is predictable.
Nevertheless, this layout may be useful when there is no need to display edges, such as node classification scenarios. 

\subsection{Layout with Similarity Matrix}
\label{sec:sm}
\begin{figure*}[tb]
	\setlength{\abovecaptionskip}{2pt}
	\setlength{\belowcaptionskip}{2pt}
	\centering
	\subfigure[]{
		\includegraphics[bb=80 50 750 700, width=0.15\textwidth]{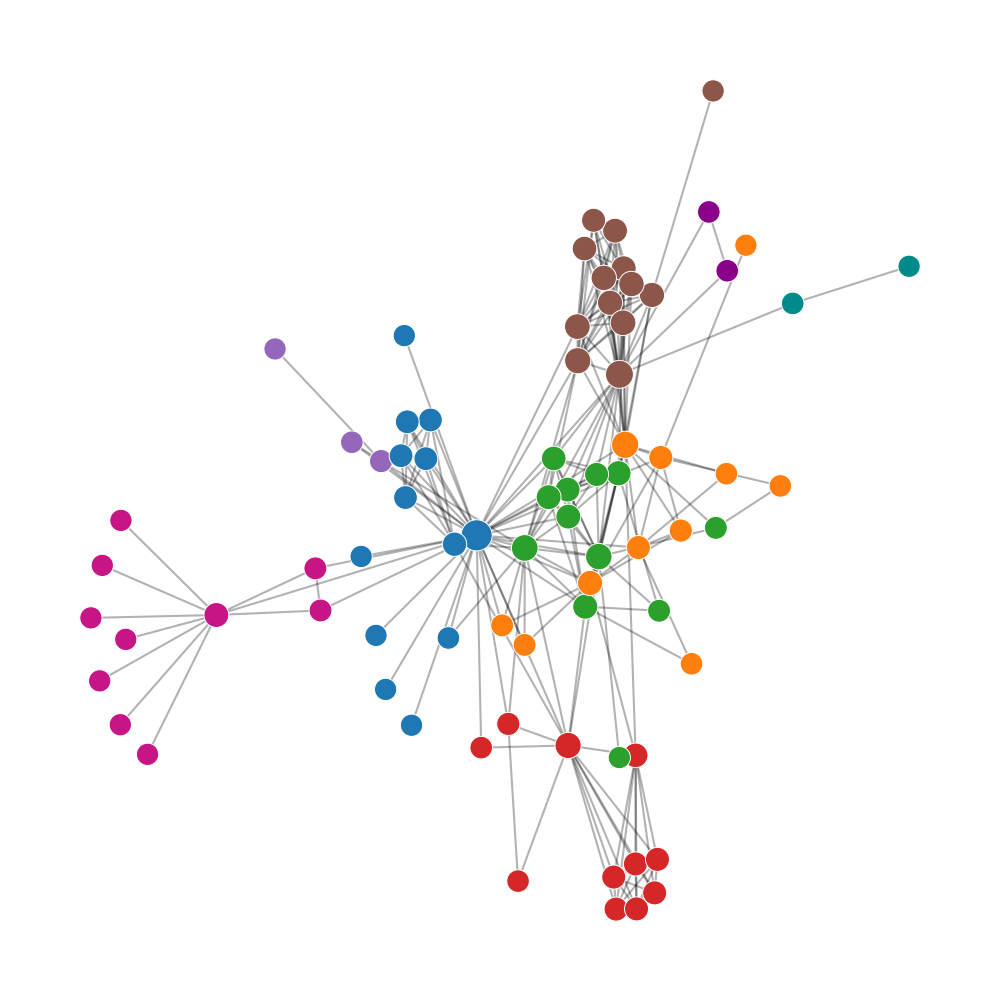}}
	\hspace{0in}
	\subfigure[]{
		\includegraphics[bb=60 50 750 700, width=0.15\textwidth]{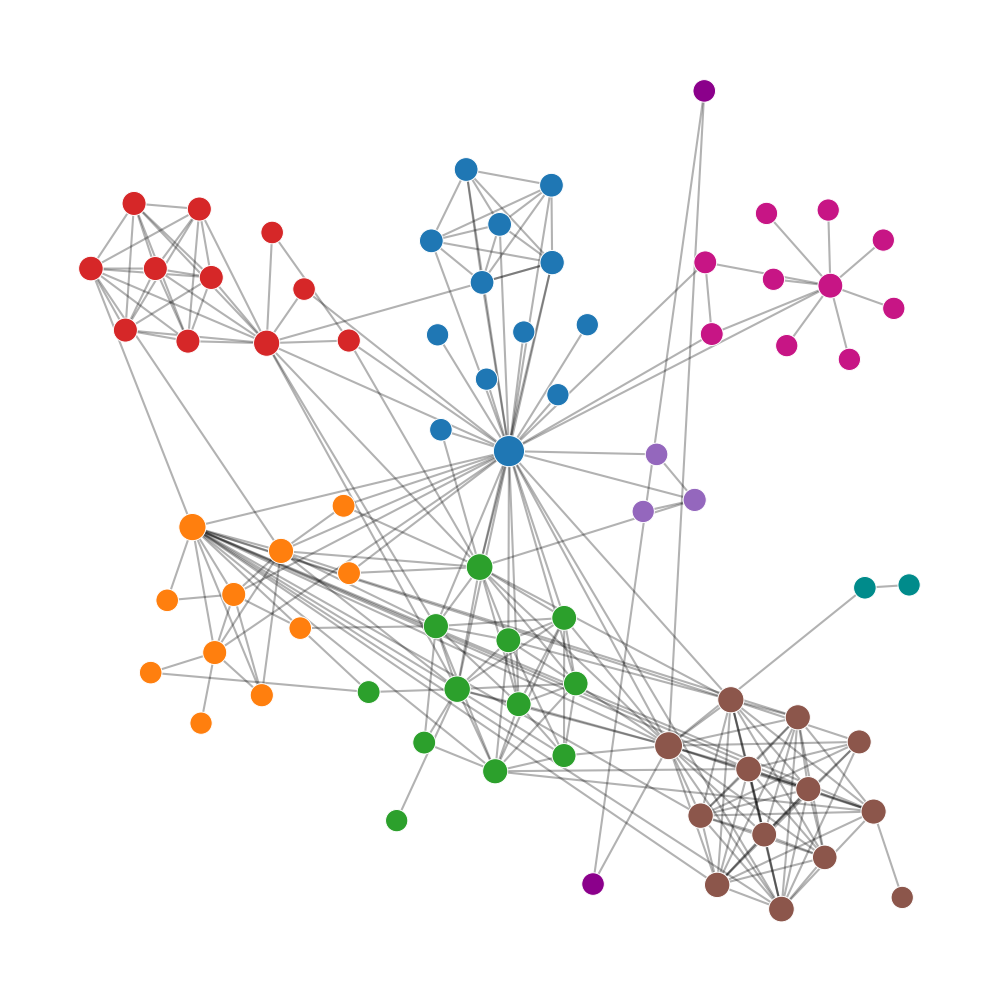}}
	\hspace{0in}
	\subfigure[]{
		\includegraphics[bb=40 50 750 700, width=0.15\textwidth]{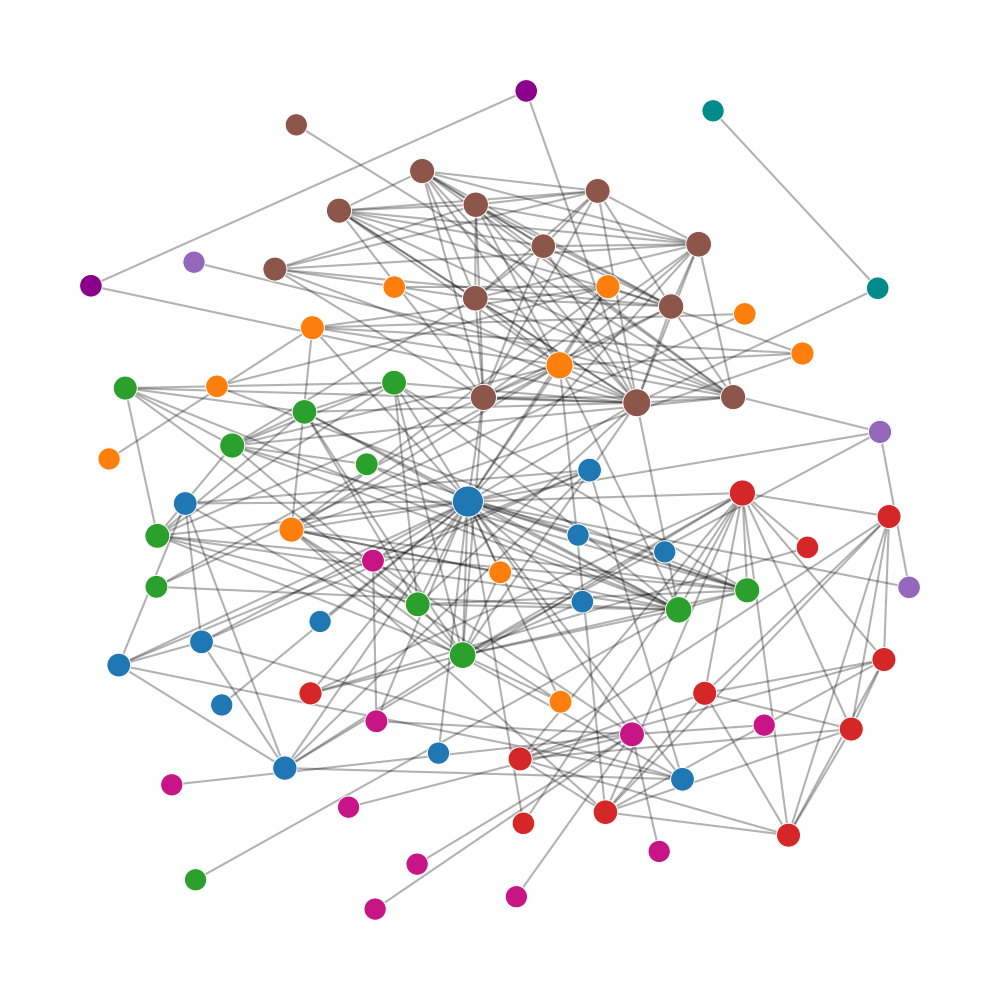}}
	\subfigure[]{
		\includegraphics[bb=80 50 750 700, width=0.15\textwidth]{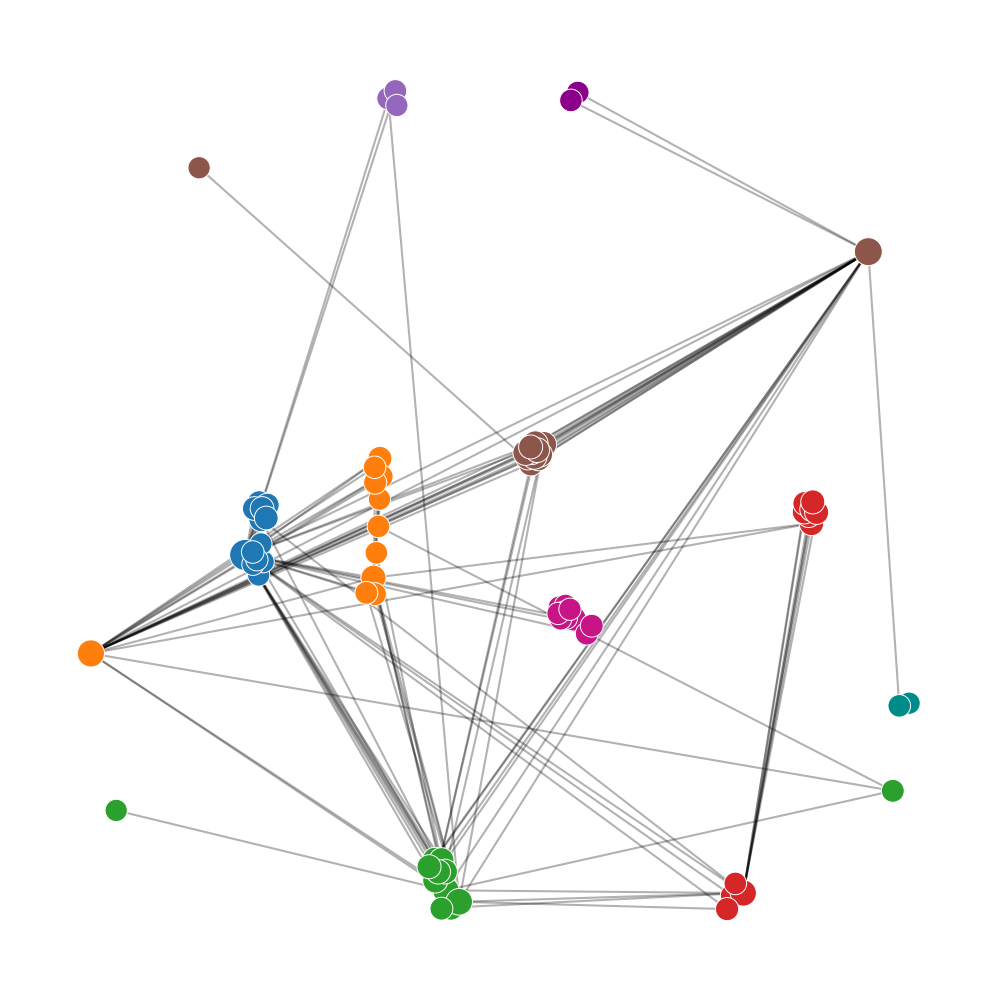}}
	\hspace{0in}
	\subfigure[]{
		\includegraphics[bb=60 50 750 700, width=0.15\textwidth]{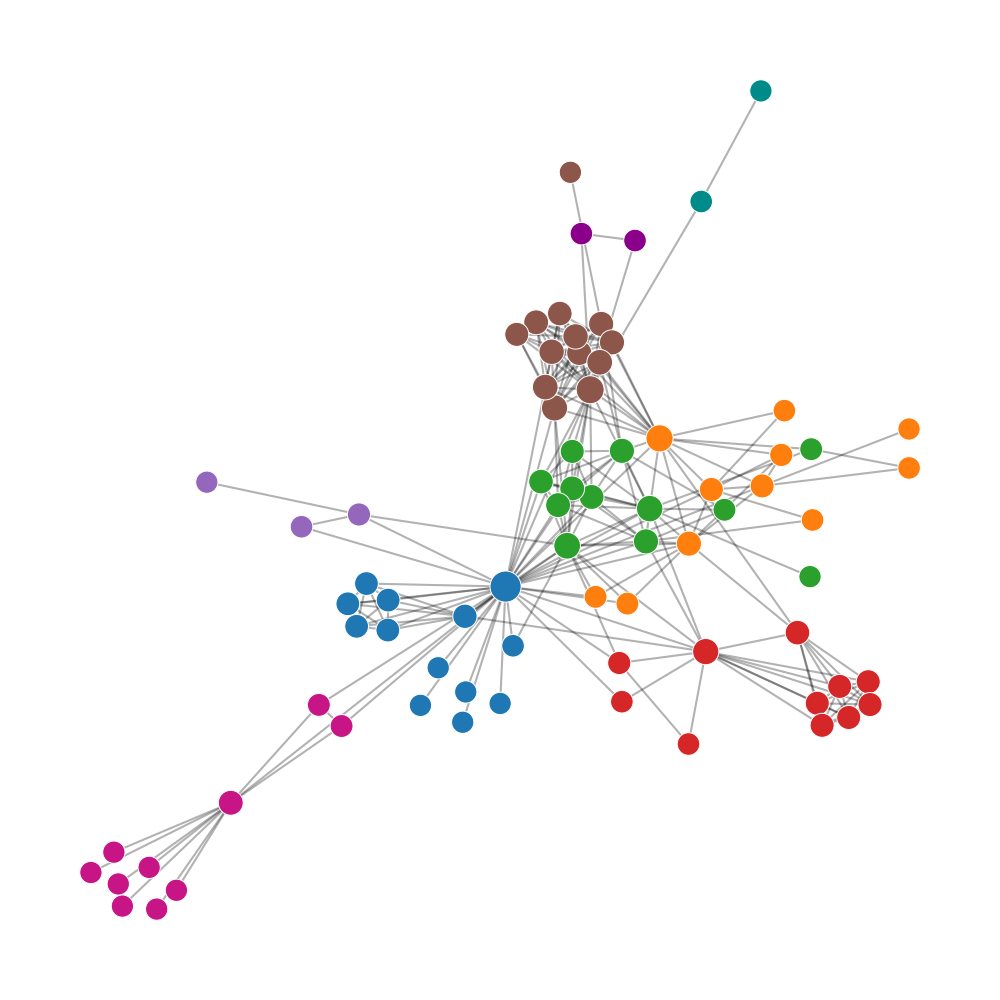}}
	\hspace{0in}
	\subfigure[]{
		\includegraphics[bb=40 50 750 700, width=0.15\textwidth]{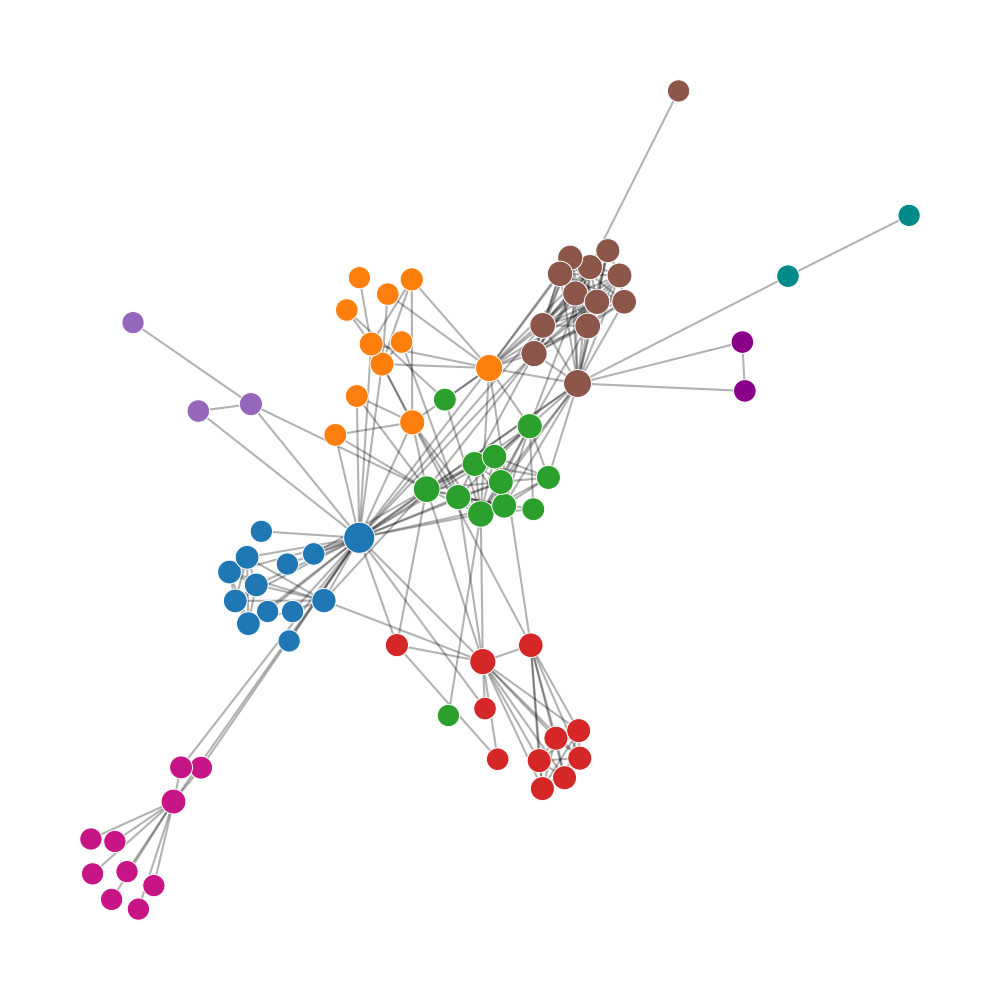}}
	\caption{
		Evolution of the embedding-driven graph layout algorithm.
		(a) is the initial layout of the Les Mis{\'e}rables dataset generated by the F-R algorithm.
		(b) is generated by directly using the dimension reduction algorithm, $t$-SNE, on the embedding vectors.
		(c) is the result of treating the similarity matrix as an adjacency matrix with F-R.
		After applying truncation on the similarity matrix, (d) is generated in the same way as (c).
		(e) is the result after integrating the similarity matrix and original adjacency matrix.
		(f) is the layout with node classification considered during the integration of
		similarity matrix and adjacency matrix.
	}
	\label{fig:basic_layout}
\end{figure*}

We later adopt the F-R algorithm to process the embedding vectors.
For each node pair in the graph, the distance between corresponding embedding vectors can be computed by euclidean distance.
So we can obtain a $n \times n$ similarity matrix $S_e$, where the values represent the similarity of two nodes.
Based on the minimum and maximum distances in the similarity matrix, each value in $S_e$ is further normalized into the range [0, 1] to generate $S_E$. Then the final similarity (normalized distances) of nodes $a$ and $b$ is formed as:
\begin{equation}
	similarity(a,b) = 1 - S_E(a,b)
\end{equation}
If $similarity(a,b)$ is larger than $similarity(a,c)$, node $a$ can be considered more similar to node $b$ than node $c$.
F-R can process graphs with or without directions and edge weights. As described in Section \ref{sec:fr}, it takes the adjacency matrix as input. 
Values in the adjacency matrix indicate the weights of edges and represent the topological connections between nodes. In fact, both the adjacency matrix and similarity matrix are shaped as $n \times n$ and indicate node connectivity and distance information. 
So we try to replace the adjacency matrix with the similarity matrix as the input of F-R.
However, as shown in Fig. \ref{fig:basic_layout} (c), the output is a hairball-like structure, and all the nodes distribute uniformly. 




\subsection{Truncation Operation on Similarity Matrix}
\label{sec:truncation}
By treating the similarity matrix as an adjacency matrix, the graph can be regarded as a new complete graph with the same nodes.
If two nodes are far from each other in the embedding space, the corresponding value in the similarity matrix is small, indicating that the two nodes are less semantic related.
Nevertheless, their impact on the simulated annealing process in F-R can be significantly magnified during the iterations, making the nodes in the layout separate uniformly. Therefore, small values in the similarity matrix should be filtered out.
To this end, we introduce a truncation function $t(x)$ to process the similarity matrix:
\begin{equation}
	t(x) = \begin{cases}
		0, & x < t_e \\
		x, & \text{otherwise}
	\end{cases}
\end{equation}
where $t_e$ is a parameter to control the threshold. 
Fig. \ref{fig:basic_layout} (d) is the layout result after applying the truncation function, where nodes in each community gather too concentrated while some others are isolated. This is because the similarity matrix generates from feature vectors, so the layout entirely relies on the embedding results. Linked node pairs with low vector similarity may be separated from each other.

\subsection{Embedding-Enhanced Adjacency Matrix}
\label{sec:integration}

\begin{figure}[t]
	\centering
	\includegraphics[width=0.46\textwidth]{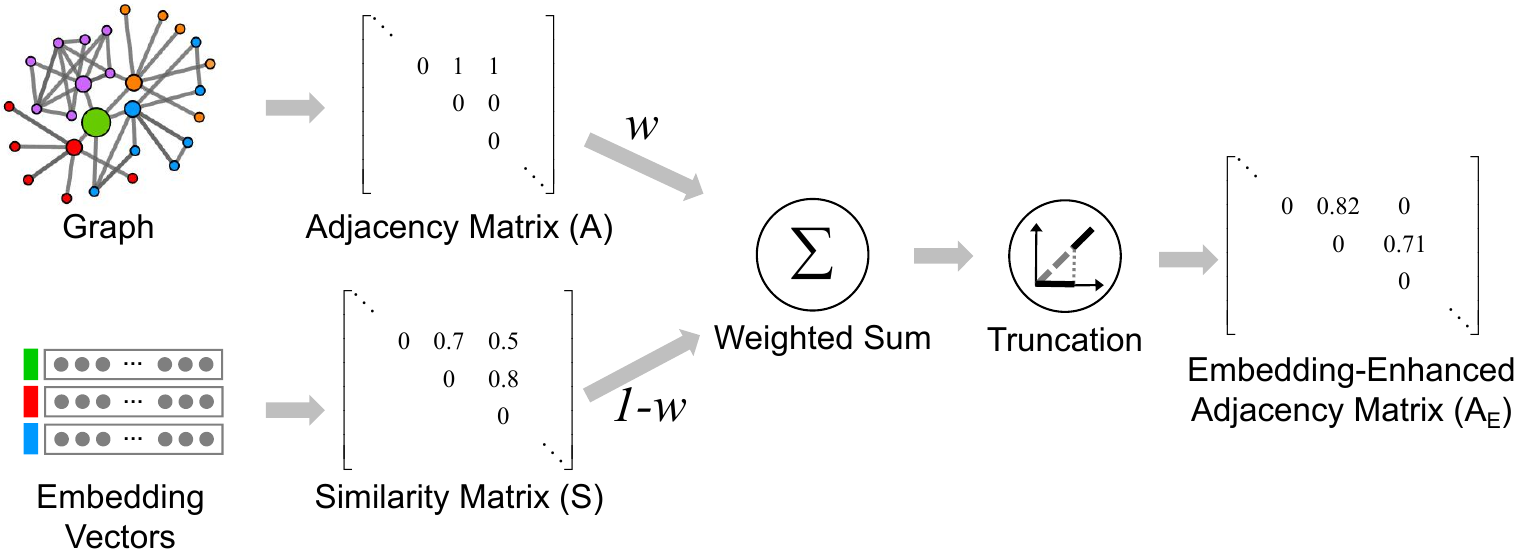}
	\caption{Embedding-enhanced adjacency matrix generation. The weighted sum of the similarity matrix generated from embedding vectors and the graph's adjacency matrix is passed to the truncation function to produce the embedding-enhanced adjacency matrix.
	}
	\label{fig:pipe}
\end{figure}

Embedding and truncation weaken the original topology information, 
which is also a major problem of dimension reduction-based methods.
Based on the attempts above, we believe that the original topology structure should be considered.
Since the adjacency matrix represents the topology structure, we propose combining the adjacency matrix and similarity matrix in the layout algorithm.
With the same size, the two matrices can be integrated by the weighted sum after normalization. So a new matrix named embedding-enhanced adjacency matrix $N$ is computed as:
\begin{equation}
N = w \times A + (1-w) \times S
\end{equation}
where $w$ is the weight parameter. $A$ and $S$ are the adjacency matrix and similarity matrix, respectively. As shown in Fig. \ref{fig:pipe}, after the weighted sum, we apply normalization and truncation on $N$ as described in Section \ref{sec:truncation} to generate the final embedding-enhanced adjacency matrix. 
By adjusting $w$, the layout result can strike a balance between Fig. \ref{fig:basic_layout} (a) and Fig. \ref{fig:basic_layout} (c). If $w < 0.5$, the similarity information from graph embedding  prevails;
if $w > 0.5$, the layout is stable, and it is a typical F-R style.
Fig. \ref{fig:basic_layout} (e) results from applying the embedding-enhanced adjacency matrix, where nodes with the same label gather into communities. In this example, the generated layout also has fewer node occlusions and edge crossings compared to the above attempts.

\subsection{Enhancement with Node Classification}
\label{sec:Node Classification}
We have achieved competitive visualizations compared to the F-R algorithm, as shown in Fig. \ref{fig:basic_layout} (e). 
Benefiting from the characteristics of graph embedding, we can further enhance the layout with classification information of nodes to produce more community-aware layouts.

With the truncation operation, we can adjust the strength of connection edges between nodes. In Sections \ref{sec:truncation} and \ref{sec:integration}, we simply apply the same truncation function to process all the nodes. However, equally treating all nodes is not conducive to presenting classification information in the layout.
To make the node clusters more distinguishable, we extend the truncation function $t(x)$ to $t'(x)$
with two additional parameters, $t_{ein}$ and $t_{eout}$, that represent the intra- and inter- cluster truncation threshold:
\begin{equation}
	t'(x, u, v) = \begin{cases}
		0, & if ~ x < p_{t}(u, v) \\
		x, & \text{otherwise}
	\end{cases}
\end{equation}
where $p_{t}$ is a function that decides which truncation threshold parameter to use:
\begin{equation}
	p_{t}(u,v) = \begin{cases}
		t_{ein}, & if ~ community(u) = community(v)  \\
		t_{eout}, & if ~ community(u) \neq community(v)
	\end{cases}
\end{equation}
where node labels provide the cluster information. We use clustering algorithms (e.g., K-Means) to cluster the nodes from embedding vectors if the label is unavailable. Generally, $t_{ein}$ is smaller than $t_{eout}$.
The final embedding-enhanced adjacency matrix $A_E$ is:
\begin{equation}
	A_E[u][v] = t'(N[u][v])
\end{equation}

\begin{figure}[t]
	\centering
	\includegraphics[width=0.46\textwidth]{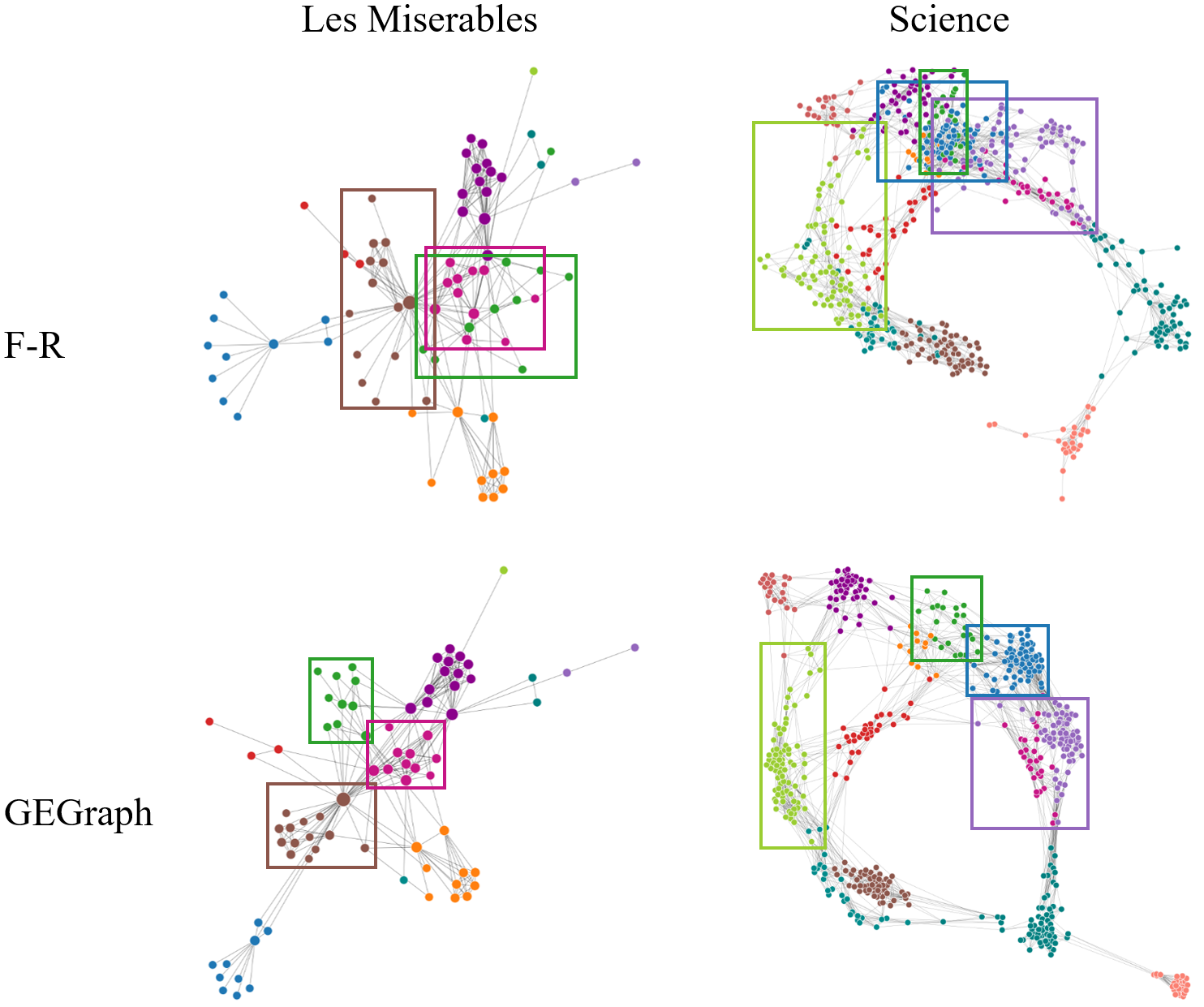}
	\caption{Detailed view of F-R and GEGraph results comparison.
	}
	\label{fig:compare_FR}
\end{figure}

Fig. \ref{fig:basic_layout} (f) is the layout result with $A_E$, and a detailed view of comparison between results of the original F-R and the improved GEGraph is shown in Fig. \ref{fig:compare_FR}.
The visualizations generated by F-R are visually appealing by taking into account a set of aesthetic criteria (e.g., distribute the verties evenly, minimize edge crossings, make edge lengths uniform, etc.), but the nodes of different communities are intertwined and difficult to discern, especially in the boxed parts of the figures. 
Compared with F-R and the above attempts, by organically embedding node connectivity and node attributes in the layout, GEGraph makes the community information more visible while retaining the original force-oriented aesthetic style of F-R. With the visualizations generated with GEGraph, we can more intuitively observe the scales of each community and how the communities link with each other.
  

The above trials reflect how we understand the connections and differences between embedding vectors and graph topology to build the final layout approach. 
In summary, as shown in Fig. \ref{fig:pipe}, first, GEGraph applies an attributed graph embedding method to generate embedding vectors from topology information and node attributes.
The vectors are used to calculate a similarity matrix, in which values represent the distances of nodes in the embedding space.
The similarity matrix reflects the high-level proximity, and the adjacency matrix represents the basic topology structure.
Weighted sum and truncation operations with clusters are used to integrate the two matrices and produce an embedding-enhanced adjacency matrix, which represents a new graph with different edge weights.
Finally, the embedding-enhanced matrix is used as the input of the F-R algorithm. 
We just take the Les Mis{\'e}rables dataset as an example here, more apparent comparisons can be found in Fig. \ref{fig:layout_result}.

\section{Embedding-Guided Exploration}\label{sec:exploration}
Various insights extracted from the embedding vectors, coupled with the embedding-driven graph layout, enable abundant graph exploration applications. In this section, we design two interactive applications as examples, including layout-preserving node aggregation and related nodes searching.

%

\subsection{Layout-Preserving Node Aggregation}
\begin{figure}[t]
	\centering
	\subfigure[Aggregation graph]{
		\includegraphics[width=0.245\textwidth]{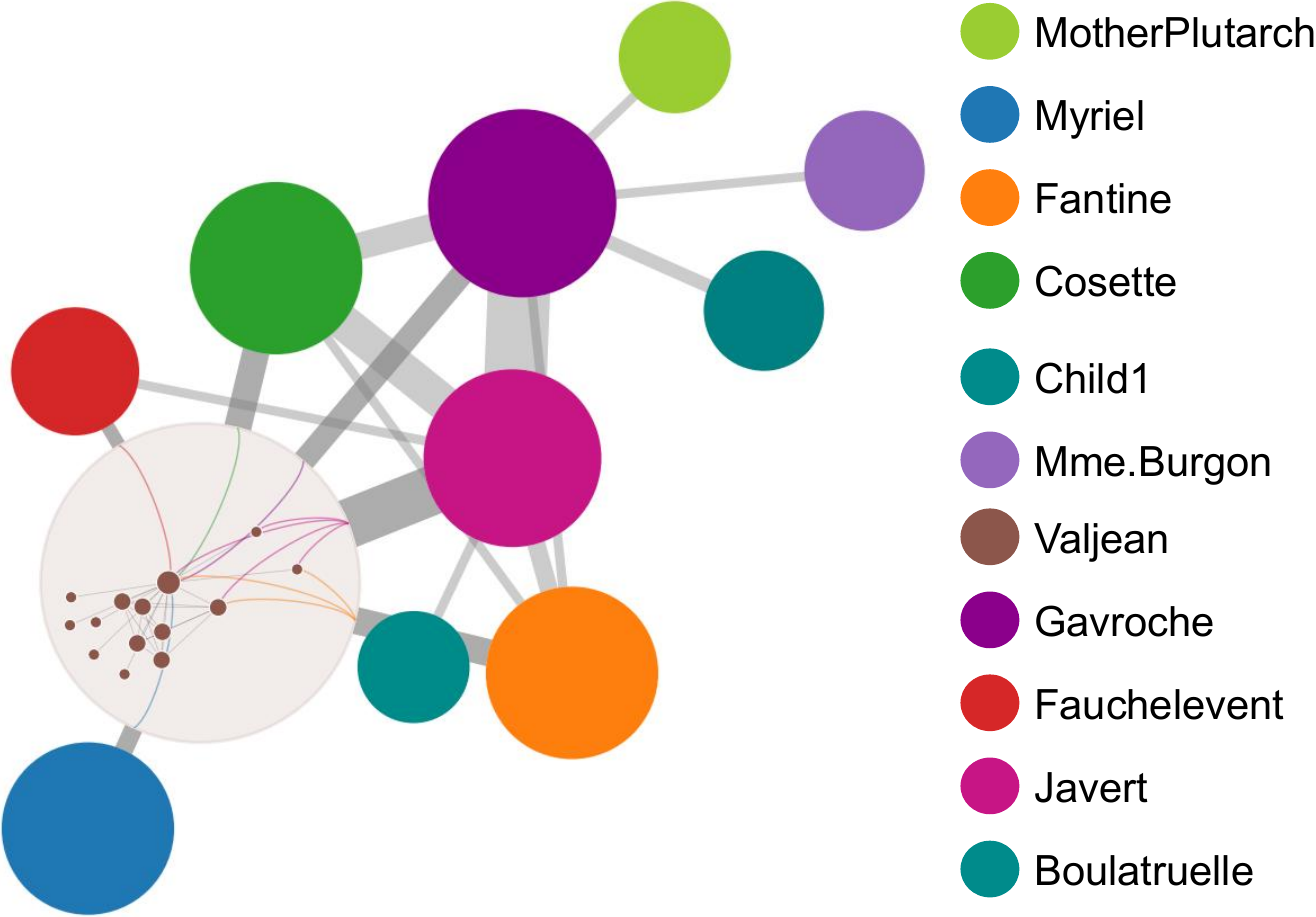}}
	\hspace{0in}
	\subfigure[Find related nodes]{
		\includegraphics[width=0.22\textwidth]{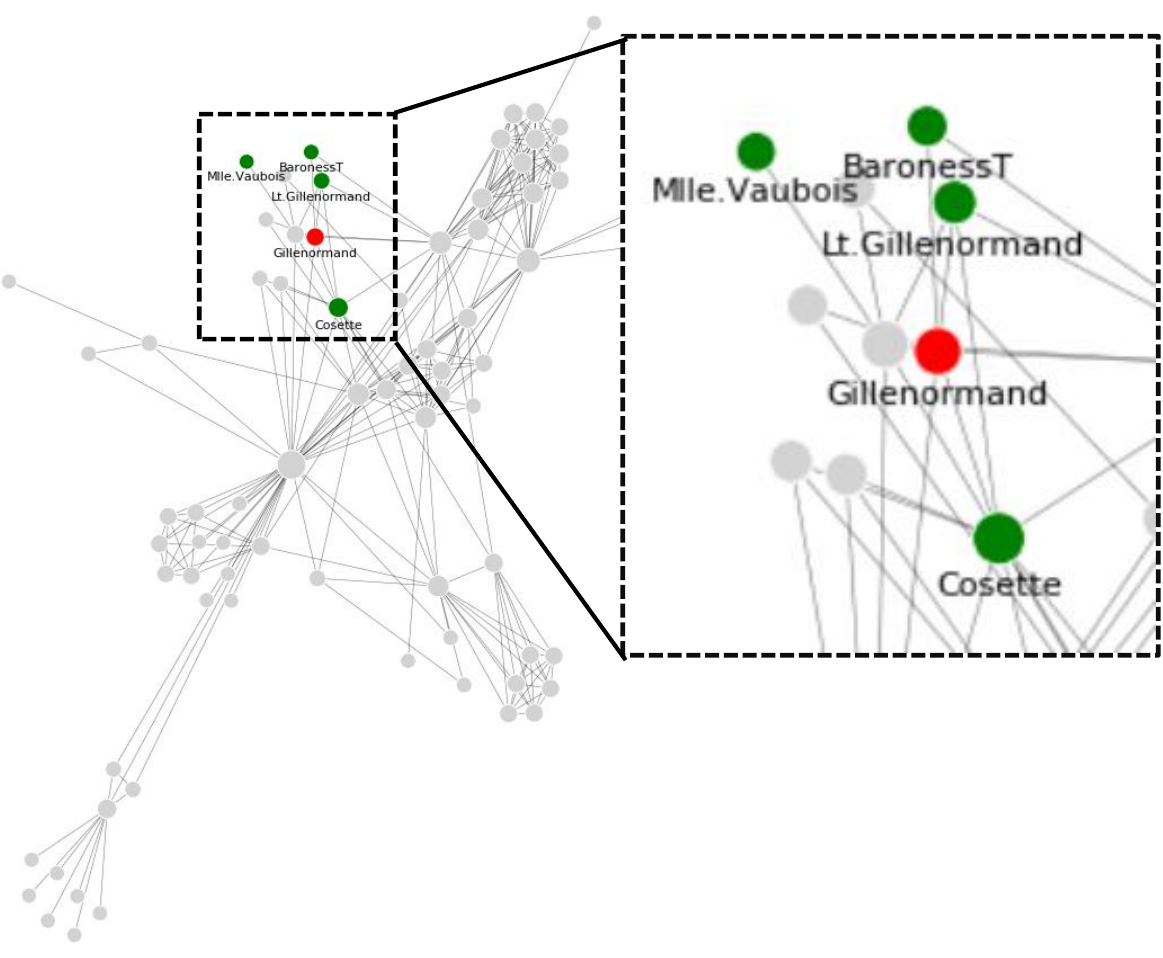}}
	
	\caption{
		Interactive exploration on the Les Mis{\'e}rables dataset based on the layout (Fig. \ref{fig:basic_layout} (f)). 
		(a) An aggregation graph to give an overview with Focus+Context interaction.
		(b) The node "Gillenormand" is clicked, and the similar nodes are stressed.
	}
	\label{fig:exploration}
\end{figure}

In graph visualization, aggregation is an effective operation to reduce the complexity of layouts and summarize the content of graphs~\cite{nobre2019state}.
In the general aggregation process, certain elements in a graph are classified, and each class is aggregated to show the connection or difference between classes~\cite{wattenberg2006visual,wongsuphasawat2017visualizing,Noack2005}.
Based on the community-aware layout (Fig. \ref{fig:basic_layout} (f)), we draw the aggregation graph (Fig.~\ref{fig:exploration} (a)) to give an overview of the original graph.
The design logic is as follows: aggregated nodes represent communities. The size of aggregated nodes reflects the scale of communities.
Two aggregated nodes are linked if an edge $(u, v)$ exists, where $u$ and $v$ are in different communities. The number of cross-community edges is reflected by the edges' width between each pair of aggregated nodes. The positions of the aggregated nodes are the centers of communities.
In addition, we compute central nodes that are representatives for communities, as shown in the legend of Fig.~\ref{fig:exploration} (a). 
The random walking paths generated in the graph embedding algorithm can be treated as ``sentences", and each node in the path is a ``word". We use TF-IDF~\cite{Robertson2004} to weight each ``word" in these ``sentences", and the node with largest numerical weight score in each group is selected as the community representative.

Inspired by Herman \textit{et al.}~\cite{herman2000graph} and responsive matrix cells~\cite{Horak2020}, we employ Focus+Context technology to facilitate interactive exploration, which allows the viewer to inspect interesting parts of the graph in detail without losing the global context. As shown in Fig.~\ref{fig:exploration} (a), the application of F+C can integrate the embedding-driven graph layout and aggregations in the same view. 
When the user clicks an aggregated node, the large aggregated node will be replaced by a subgraph constructed with the real nodes belonging to the community.
Additionally, 
inspired by an edge bundling work of Wang \textit{et al.} ~\cite{wang2019interactive},
we design the transition effect from aggregated edges to focused layout edges.
The aggregated edge terminates on the boundary circle, and the cross-community edges are linked from the endpoint on the circle to real nodes.
In order to emphasize that these edges link the two communities, the Bezier curve algorithm is applied to make their shapes better.
These edges are also drawn with different colors to indicate which community they link to.
As a result, users can obtain both the global layout and the detailed community structure with an engaging user experience.

\subsection{Multi-Strategy Related Nodes Searching}

Another embedding-driven exploration method we design is to interactively find the related nodes of a specific node, which is a common visual exploration task in the graph~\cite{lee2006}.
Existing solutions usually display neighbors with first-order or second-order proximity based on the graph topology. Despite the simplicity and intuitiveness, they cannot reveal the comprehensive similarity or higher-order proximity of nodes.

In our application design, after graph embedding, the nodes are encoded as feature vectors, and the similarity of the nodes can be calculated with different proximities, as described in Section \ref{sec:strategy}.
For one specific node, 
with different random walk strategies, the layout can find related nodes in three embedding spaces: local structure level, global structure level, and attribute level.
For example, in a citation graph of the Science dataset~\cite{suh2019persistent}, the local proximity of nodes implies the direct citation between papers. The attribute proximity can find the papers in the same community. The global proximity indicates the similar structure role in different communities.
As shown in Fig.~\ref{fig:exploration} (b), by clicking one node in the graph, the similar nodes will be labeled and stressed with green color through appropriate searching strategies with vector distances.
As a result, users can search related nodes in the graph with different proximities to satisfy their various exploration tasks.

\begin{figure*}[b]
	\setlength{\abovecaptionskip}{2pt}
	\setlength{\belowcaptionskip}{2pt}
	\centering
	\includegraphics[width=0.98\textwidth]{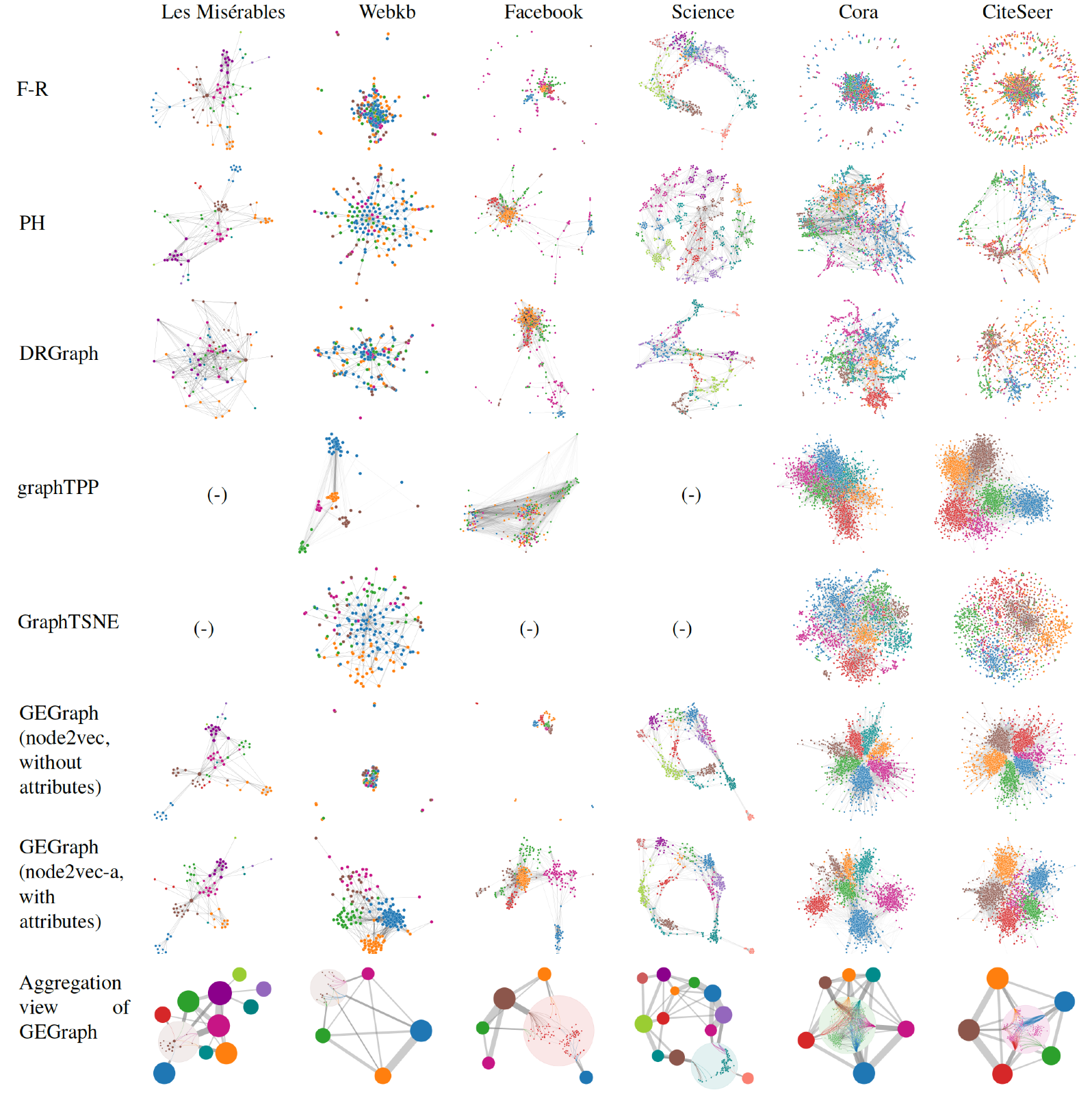}

	\caption{Graph visualizations with F-R, PH, DRGraph, GraphTSNE, graphTPP, and GEGraph, as well as the aggregation view.
	}
	\label{fig:layout_result}
\end{figure*}

\begin{table}[t]
	\scriptsize
	\caption{The graph datasets used in the evaluation.}
	\label{tab:graph_datasets}
	\centering
	\setlength{\tabcolsep}{1.8mm}{
		\begin{tabular}{m{2.09cm}m{0.41cm}m{0.4cm}m{0.76cm}m{0.39cm}m{2.59cm}}
			\hline
			\rowcolor[HTML]{EFEFEF} 
			Name           & \#node & \#edge & \#attribute & \#label & Description             \\ 
			\toprule
			Les Mis{\'e}rables~\cite{knuth1993stanford} & 77 & 254 & 0 & Yes & Characters network of Victor Hugo’s novel. \\ \hline
			Webkb(Cornell)~\cite{craven1998learning} & 195 & 286 & 1703 & Yes & Webpage citation network of Cornell University \\ \hline
			Facebook~\cite{leskovec2012learning} & 347 & 2519 & 224 & No & Social network from Facebook. \\ \hline
			Science~\cite{borner2012design} & 554 & 2276 & 0 & Yes & Cross-disciplinary coauthorship in science. \\ \hline
			Cora~\cite{lu2003link} & 2708 & 5278 & 1433 & Yes & Citation network of scientific publications. \\ \hline
			CiteSeer~\cite{lu2003link} & 3264 & 4536 & 3703 & Yes & Citation network of scientific publications.    \\
			\bottomrule
	\end{tabular}}
\end{table}

\section{Evaluation}
\label{sec:evaluation}
We evaluate our approach from four perspectives: 
(1) We validate the layout quality qualitatively with multiple graph datasets; 
(2) We adopt a set of quantitative metrics to measure the readability of graph layouts; 
(3) We conduct a user study for human-centered analysis;
(4) We present two case studies on real-world data.

\subsection{Experiment Setting}\label{sec:setting}
The settings include implementation, datasets, baselines to compare, and the optional selection of parameters.

\textbf{Implementation. }
We coded our approach with Python and referred to the F-R code of Networkx\footnote{https://github.com/networkx/networkx} and an efficiency-optimized version of node2vec\footnote{https://github.com/VHRanger/nodevectors}. Our code\footnote{https://github.com/tzw28/EmbeddingGuidedLayout} was run in multiple threads on a PC with Intel(R) Core(TM) i7-9700 CPU, and 16 GB memory.

\textbf{Datasets. }
Table~\ref{tab:graph_datasets} lists the information of graph datasets used in the evaluation, including the number of nodes, edges, attributes, and whether the class label is provided. Nodes in Les Mis{\'e}rables and Science are not attached with node attributes, and the label of nodes is used as the only attribute during graph embedding.

\textbf{Baseline.}
We compare our layout algorithm, GEGraph, with five representative methods (F-R, PH, DRGraph, graphTPP, and GraphTSNE). Our selection prioritizes approaches that have a wide range of applications, a publicly available implementation, and a record of comparison over others. 
The selected methods cover two classification dimensions. For groupings from Nobre \textit{et al.}~\cite{nobre2019state}, F-R, PH, and DRGraph are topology-driven, graphTPP is attribute-driven, while GraphTSNE is a hybrid approach. For families in ~\cite{gibson2013survey,VonLandesberger2011}, DRGraph and GraphTSNE are dimensionality reduction-based, F-R, PH, and graphTPP are force-directed, while DRGraph is a multi-level approach.
In addition, to explore the importance of integrating attributes in the layout, we also include a node2vec-based version that only embeds the topology information, GEGraph (node2vec). GEGraph (node2vec-a) is the final version of integrating attributes.

\textbf{Selection of Parameters.} There are several pending tunable parameters in our design. Similar to DRGraph~\cite{Zhu2021}, we intensively discuss the selection of parameters here. We have set default values for simplicity while achieving relatively good results based the parameter sensitivity analysis. Users are also allowed to finetune the parameters according to their requirements. 
\begin{itemize}
\item 
\textit{Random-walk strategy parameters $p$, $q$, and $r$}. 
As demonstrated in Section \ref{sec:strategy}, users can adjust $p$, $q$, and $r$ to control the probability of choosing next node during random walk process. For the purpose of drawing neighbors or attribute-similar nodes closer, we often use a lower value ($ < 1$) of $q$ and $r$ and simply set $p = 1$ in such cases. 

\item 
\textit{The matrix weight $w$}. 
As described in Section \ref{sec:integration}, $w$ balances the embedding features and topological information. Generally, we set $w = 0.4$ to make embedding information more significant while keeping the overall topology of the graph in the layout based on the sensitivity analysis.

\item 
\textit{The truncation threshold $t_{ein}$ and $t_{eout}$}. 
As discussed in Section \ref{sec:Node Classification}, the embedding-enhanced adjacency matrix is filtered by the truncation function with the threshold $t_{ein}$ and $t_{eout}$. A high $t_{ein}$ value will weaken connections between nodes in the same cluster and a low $t_{eout}$ value will keep most connections between clusters. The layout is more sensitive to $t_{ein}$ than $t_{eout}$. We use $t_{ein} = 0.4$ and $t_{eout} = 0.6$ in general cases based on the sensitivity analysis.

\end{itemize}

In addition, the number of iterations of the F-R algorithm is also a parameter. We are consistent with the design of the original F-R algorithm, which defaults to 50 iterations. As discussed in~\cite{fruchterman1991graph}, although this is excessive on the smaller graphs, the iterations of small graphs consume little time. So in general users do not need to adjust it. For very large graphs, users can increase the number of iterations appropriately.

\subsection{Qualitative Layout Quality}
\label{sec:visualization_results}
Fig.~\ref{fig:layout_result} shows the visualizations generated by GEGraph and baseline methods.
Generally, GEGraph avoids manual adjustments and prior knowledge requirement and allows more input graph types. We can observe evident community information and fewer node occlusions and edge crossings in the layouts created by GEGraph (node2vec-a) in the study compared to the other methods under evaluation.
In detail, F-R cannot handle the scale properly when dealing with unconnected graphs, such as Webkb, Facebook, Cora, and CiteSeer, where the discrete nodes are far from the connected body. When dealing with large graphs, it usually produces poor layouts as it converges to local minima and can hardly retain the global structure.
PH initially draws a graph with F-R. Then the user can click the persistent barcodes to separate the partitions generated by persistent homology features. 
Its bar-list integration design is novel but not effective enough for large graphs. On Cora and CiteSeer, we fail to identify the most compelling portion barcodes among thousands of them and spend considerable time drawing a relatively good layout with PH.
DRGraph focuses on efficiently processing large graphs. So it performs relatively well on large datasets (e.g., Cora and CiteSeer). However, it can hardly achieve good global layouts, and the clutter of unrelated structures usually masks the community information. 
The above three methods are all topology-driven and ignore the node attributes.
Yet, graphTPP uses the attribute-based cluster information to draw community-aware visualizations. However, it ignores the typological structure. So the generated layout usually contains compact clusters, where most nodes are positioned at the same coordinates.
GraphTSNE can integrate node attributes with topology. It performs slightly better than topology-driven approaches, but the community information is not distinguishable enough compared with GEGraph. As GraphTSNE can only process graphs with multiple attributes, Les Mis{\'e}rables, Facebook, and Science are skipped.
GEGraph integrates node attributes with topological information while preserving F-R's "hub-and-spoke" drawing style. 
The communities are clarified on small graphs like Les Mis{\'e}rables and Webkb, while the topological structure still plays a primary role. On large graphs like Cora and CiteSeer, GEGraph can separate nodes into distinct communities to avoid a messy visualization. The aggregation views also indicate the community awareness of GEGraph. 
If without attributes embedded, the layouts generated by GEGraph (node2vec) are relatively poor, which is evident in Webkb and Facebook. 
So our method can produce attractive and community-aware graph layouts, striking a better balance between aesthetic and exploration goals.


\subsection{Quantitative Measurement}
We further measure the layout results in Fig. \ref{fig:layout_result} with a set of quantitative readability metrics.

\subsubsection{Evaluation Metrics}
We adopt six metrics (three for aesthetic goals and three for exploration goals) derived from the evaluation frameworks proposed by Haleem \textit{et al.}~\cite{haleem2019evaluating} and Wang \textit{et al.}~\cite{Wang2016d}. These metrics are also widely used to evaluate other graph layout algorithms, which are introduced as follows:


\begin{itemize}[leftmargin=*]
	\item
	 \textbf{Node Spread ($N_{sp}$)} evaluates the node dispersion that measures the average distance of nodes from the community center.
	\begin{equation}
		N_{sp} = \sum_{c\in C}\frac{1}{|C|}\sum_{v\in c}\sqrt{(v_x-c_x)^2+(v_y-c_y)^2}
	\end{equation}
	where $C$ is the set of communities, $(v_x, v_y)$ and $(c_x, c_y)$ are the position of the node and community center, respectively. 
	Higher $N_{sp}$ indicates less distinguishable communities in the layout.
	
	\item
	\textbf{Node Occlusions ($N_{oc}$)} is the count of node pairs that are placed at the same coordinate within a threshold. The count is further divided by total number of pairs.

	\item
	\textbf{Edge Crossings ($E_{c}$)} evaluates the frequency of edge crossing, which can cause clutter in the layout. General edge crossings ($E_{c}$) is the percentage of crossed edges among all edge pairs. 

	\item
	\textbf{Group Overlap ($G_{o}$)} measures the overlap between communities' convex hulls, and a small value indicates the visually apparent group membership in the graph.
	\begin{equation}
		G_{o} = 1-\frac{1}{|P|}\sum_{g\in P}\frac{overlap(g, P \setminus g)}{|P \setminus g|}
	\end{equation}
	where $overlap(g, P \setminus g)$ is the number of nodes within the convex hull between the community $g$ and other communities $P \setminus g$.
	
	\item
	\textbf{Community Entropy ($H$)} measures the disorder of the nodes in the layout. The layout area is divided into uniform rectangular regions. For each region, the nodes in the region belong to different communities. $p(c)$ is the percentage of nodes from community $c$, then the entropy of the local region is:
	\begin{equation}
		H = -\sum_{c\in C}p(c)\log_{2}{p(c)}
	\end{equation}

	\item
	\textbf{Spatial Autocorrelation ($C$)} is a geographical data metric that measures the community distribution of the layout. For the region within a certain radius around the node $i$, $C_{i}$ is calculated as:
	\begin{equation}
		C_{i} = \frac{\sum_{j=1}^{N}(1-NormDist(i,j)) IsSameCommunity(i, j)}{\sum_{j=1}^{N}(1-NormDist(i,j))}
	\end{equation}
	where $N$ is number of nodes around node $i$, $NormDist(i, j)$ is the normalized distance between the two nodes $i$ and $j$. If they are from the same community, $IsSameCommunity(i, j)$ returns $0$, otherwise $1$. $C$ is the average value of all $C_{i}$.
	 
\end{itemize}

\begin{table}[t]
	\scriptsize
	\caption{Quantitative evaluation results. Smaller values indicate better performance in specific metrics. Outliers are marked red, and values of the best performance are highlighted in bold font.}
	\setlength{\tabcolsep}{1.9mm}{
	\begin{tabular}{m{1.2cm}m{2.4cm}m{0.35cm}m{0.35cm}m{0.35cm}m{0.35cm}m{0.35cm}m{0.35cm}}
		\hline
		\rowcolor[HTML]{EFEFEF} 
	       Dataset & Method & $N_{sp}$ & $N_{oc}$ & $E_{c}$ & $G_{o}$ & $H$ & $C$ \\ \hline
	\multirow{7}*{Les Mis{\'e}rables} & F-R & 0.076 & \textbf{0.000} & 0.027 & 0.009 & 0.500 & 0.343 \\ 
	~ & PH & \textbf{0.051} & \textbf{0.000} & \textbf{0.025} & 0.002 & 0.302 & 0.166 \\ 
	~ & graphTPP & - & - & - & - & - & - \\
	~ & DRGraph & {\color[HTML]{c95862} \textbf{0.150}} & 0.068 & {\color[HTML]{c95862} \textbf{0.119}} & {\color[HTML]{c95862} \textbf{0.189}} & {\color[HTML]{c95862} \textbf{1.043}} & {\color[HTML]{c95862} \textbf{0.732}} \\ 
	~ & GraphTSNE & - & - & - & - & - & - \\ 
	~ & GEGraph(node2vec) & 0.058 & \textbf{0.000} & 0.027 & \textbf{0.000} & 0.220 & 0.217 \\ 
	~ & \textbf{GEGraph(node2vec-a)} & \textbf{0.051} & \textbf{0.000} & 0.028 & \textbf{0.000} & \textbf{0.174} & \textbf{0.134} \\ \hline
	
	\multirow{7}*{Webkb} & F-R & 0.158 & 0.158 & 0.011 & 0.411 & 0.775 & 0.707 \\ 
	~ & PH & 0.256 & 0.067 & \textbf{0.005} & 0.326 & 1.118 & 0.667 \\
	~ & graphTPP & \textbf{0.038} & {\color[HTML]{c95862} \textbf{3.508}} & {\color[HTML]{c95862} \textbf{0.214}} & \textbf{0.000} & \textbf{0.034} & \textbf{0.015}\\
	~ & DRGraph & 0.232 &0.310&0.006&0.332&1.174&0.707\\
	~ & GraphTSNE & 0.265 & 0.095 & 0.028 & 0.185 & 1.116 & 0.488 \\
	~ & GEGraph(node2vec) & 0.084 & {\color[HTML]{c95862} \textbf{5.396}} & 0.014 & 0.385 & 0.806 & 0.719 \\
	~ & \textbf{GEGraph(node2vec-a)} & 0.109 & \textbf{0.021} & 0.069 & 0.007 & 0.268 & 0.143 \\ \hline
	
	\multirow{7}*{Facebook} & F-R & - & {\color[HTML]{c95862} \textbf{2.061}} & 0.068 & - & - & - \\
	~ & PH & - & 0.820 & 0.066 & - & - & - \\
	~ & graphTPP & - & {\color[HTML]{c95862} \textbf{1.693}} & {\color[HTML]{c95862} \textbf{0.192}} & - & - & - \\
	~ & DRGraph & - & 0.493 & 0.088 & - & - & - \\
	~ & GraphTSNE & - & - & - & - & - & - \\
	~ & GEGraph(node2vec) & - & {\color[HTML]{c95862} \textbf{6.732}} & 0.090 & - & - & - \\
	~ & \textbf{GEGraph(node2vec-a)} & - & \textbf{0.120} & \textbf{0.060} & - & - & - \\ \hline
	
	\multirow{7}*{Science} & F-R & 0.075 & 0.103 & \textbf{0.003} & 0.018 & 0.661 & {\color[HTML]{c95862} \textbf{0.386}} \\
	~ & PH & 0.098 & \textbf{0.000} & 0.004 & 0.007 & 0.502 & \textbf{0.171} \\
	~ & graphTPP & - & - & - & - & - & - \\
	~ & DRGraph & 0.083 & {\color[HTML]{c95862} \textbf{0.355}} & \textbf{0.003} & 0.033 & 0.676 & 0.285 \\
	~ & GraphTSNE & - & - & - & - & - & - \\
	~ & GEGraph(node2vec) & \textbf{0.051} & 0.145 & \textbf{0.003} & 0.003 & 0.566 & 0.275 \\
	~ & \textbf{GEGraph(node2vec-a)} & \textbf{0.051} & 0.147 & \textbf{0.003} & \textbf{0.002} & \textbf{0.458} & 0.229 \\ \hline
	
	\multirow{7}*{Cora} & F-R & 0.099 & {\color[HTML]{c95862} \textbf{0.571}} & 0.004 & {\color[HTML]{c95862} \textbf{0.419}} & 1.268 & 0.751 \\
	~ & PH & 0.206 & {\color[HTML]{c95862} \textbf{0.407}} & 0.004 & 0.232 & 1.210 & 0.488 \\
	~ & graphTPP & 0.098 & 0.126 & 0.027 & 0.101 & 0.548 & 0.386 \\
	~ & DRGraph & 0.155 & 0.207 & \textbf{0.002} & 0.209 & 1.043 & 0.388 \\
	~ & GraphTSNE & 0.184 & \textbf{0.067} & 0.004 & 0.202 & 0.934 & 0.319 \\
	~ & GEGraph(node2vec) & 0.115 & 0.120 & 0.019 & 0.025 & 0.550 & 0.254 \\
	~ & \textbf{GEGraph(node2vec-a) }& \textbf{0.092} & 0.131 & 0.014 & \textbf{0.009} & \textbf{0.402} & \textbf{0.111} \\ \hline
	
	\multirow{7}*{Citeseer} & F-R & 0.202 & 0.299 & 0.003 & 0.446 & {\color[HTML]{c95862} \textbf{2.041}} & 0.770 \\
	~ & PH & 0.234 & {\color[HTML]{c95862} \textbf{1.347}} & 0.004 & 0.347 & 1.271 & 0.507 \\
	~ & graphTPP & 0.095 & 0.087 & 0.029 & 0.028 & 0.444 & 0.148 \\
	~ & DRGraph & 0.241 & 0.258 & \textbf{0.001} & 0.422 & 1.580 & 0.590 \\
	~ & GraphTSNE & 0.203 & \textbf{0.079} & 0.002 & 0.319 & 1.197 & 0.452 \\
	~ & GEGraph(node2vec) & 0.091 & 0.191 & 0.028 & 0.005 & 0.328 & 0.163 \\
	~ & \textbf{GEGraph(node2vec-a)} & \textbf{0.079} & 0.158 & 0.014 & \textbf{0.004} & \textbf{0.220} & \textbf{0.044} \\ \hline
	\end{tabular}}
	\label{tab:quantitative}
		\vspace{-0.2cm}
\end{table}

\subsubsection{Quantitative Results}
The quantitative evaluation results of the layouts in Fig. \ref{fig:layout_result} are listed in Table~\ref{tab:quantitative}. Smaller values for all metrics indicate better layout quality, and the magnitude of values is related to the dataset scale. 
In each group with a specific dataset and a metric, we mark the outlier values of extreme poor performance in red that lead to confusing layouts, mainly based on the outlier detection algorithm in the boxplot. Values of best performance in each group are highlighted in bold font. 
Since our goal is to strike a better balance between aesthetic and exploration goals, we expect a good visualization to perform well on both sets of metrics, at least well on one dimension but also above average on the other.
Some algorithms can achieve the best score in a set of metrics but have outlier valves in other metrics, so the resulting layout is still confusing or not community-aware.

\textbf{Exploration-related metrics} ($G_{o}$, $H$, and $C$):
The metrics measure community awareness of the layouts, which is vital for graph exploration. Compared with other methods, it is evident that GEGraph (node2vec-a) achieves the best performance in most communities-based cases. An exception is graphTPP with the Webkb dataset. It clusters nodes into visible communities, so the corresponding community-related metrics are small. However, the nodes in the community are too dense (Fig. \ref{fig:layout_result}), leading to outlier metric values of node occlusion and edge crossing, which is mainly due to the ignorance of the graph topology. Similar cases are also seen with F-R, PH, and the node2vec-based GEGraph (node2vec). Without attributes considered, they mostly achieve relatively high values in exploration-related metrics and can hardly produce community-aware visualizations.
The visualizations generated by GraphTSNE do not strike a good balance between aesthetic and exploration goals, even though they take into account the graph connectivity and node attributes. As shown in Fig. \ref{fig:layout_result}, the nodes are evenly distributed compared to GEGraph (node2vec-a)'s results, and the community information is not obvious.

\textbf{Aesthetics-related metrics} ($N_{sp}$, $N_{oc}$, and $E_{c}$):
GEGraph (node2vec-a) is superior to others in terms of $N_{sp}$ because the introduction of attributed graph embedding makes related nodes gather closely and the communities are distinguishable. 
GEGraph (node2vec-a) performs well on $N_{oc}$ and $E_{c}$ with small datasets (e.g., Facebook) and has a not bad performance with large datasets (at least there are no outliers).
So the layouts' community awareness comes at the cost of a small increase in node occlusions and edge crossings on top of the visualizations remaining visually appealing.
F-R, PH, graphTPP, and DRGraph include outliers in terms of $N_{oc}$ with specific datasets, which lead to the nodes placed in a compact form.
GraphTSNE has no outliers on the metrics but most values are at an intermediate level, so the communities are not visually distinguishable enough, and there is still some clutter in the layout. 

Overall, GEGraph outperformed on the metrics, especially on the community-related metrics.
The generated visualizations can highlight community information while retaining the aesthetics of the layout.

\begin{figure}[t]
	\setlength{\abovecaptionskip}{2pt}
	\setlength{\belowcaptionskip}{2pt}
	\subfigure[Distribution of participants' selection]{
		\includegraphics[width=0.4\textwidth]{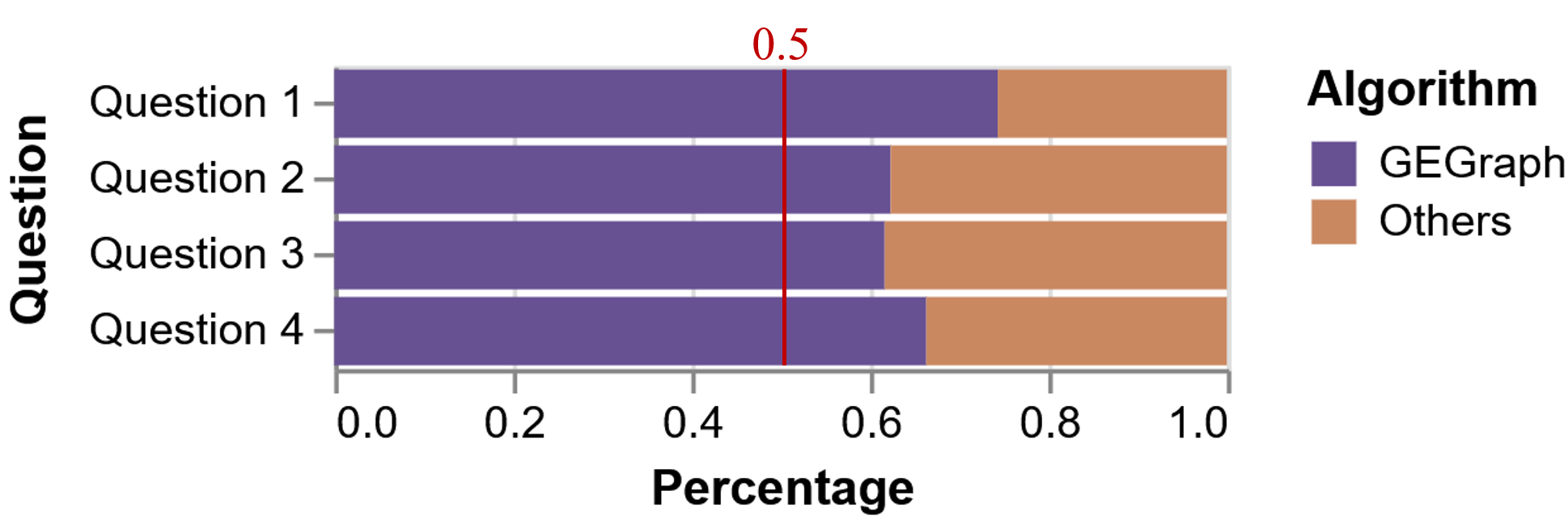}
		\label{fig:study 1}	
	}
	
	\subfigure[User ratings of exploration applications with a 5-point Likert scale]{
		\includegraphics[width=0.48\textwidth]{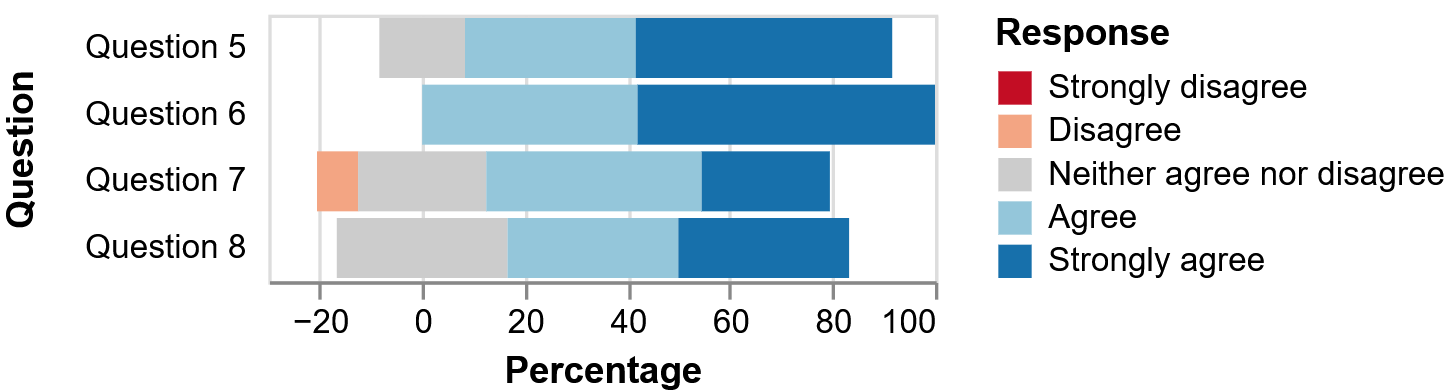}
		\label{fig:study 2}	
	}	
	\caption{
		Results of the user study.
	}
	
	\vspace{-0.3cm}
\end{figure}

\subsection{User Study}\label{sec:user study}	
We conducted a user study for human-centered evaluation. We recruited 12 participants (5 females and 7 males) who had experience in data analysis, covering university students, data analysts, and software engineers. We reorganized the visualizations of each dataset in Fig. \ref{fig:layout_result} by pairing the results from GEGraph and each of the other methods. In the study, we first explained the underlying graph data and then showed participants the layout pairs (the order was random) through an questionnaire. For each visualization, users can zoom in to observe the graph structure and semantics.
Finally, we asked them to compare the listed visualizations and answer the following questions on the questionnaire.
	
	\noindent (1)
	Which visualization do you think is easier to explore clusters?
	
	\noindent (2)
	Which visualization is more visually appealing?
	
	\noindent (3)
	Which visualization better helps you understand the graph?
	
	\noindent (4)
	Which visualization will you choose for further analysis?
	
After that, we showed the two applications (node aggregation and related nodes searching) and asked the participants to freely try with graph datasets in Table \ref{tab:graph_datasets}. 
They were allowed to use a mouse to interact with the graph on a laptop as described in Section \ref{sec:exploration} and then answer the following 5-point Likert scale questions:
	
	\noindent (5)
	Node aggregation can give a good overview of the underlying graph layout.
	
	\noindent (6)
	The design of node aggregation can help me explore graphs.
	
	\noindent (7)
	Related nodes searching can give appropriate results based on the three strategies.
	
	\noindent (8)
	Related nodes searching can help me explore graphs.
	
	Fig. \ref{fig:study 1} shows the distribution of participants’ selection of better visualizations. The results indicate that the participants rated GEGraph better in most pairs, considering community preservation, aesthetics, and exploration goals. 
	Several participants commented that they made a decision quickly and found the results of GEGraph better in most cases. Some also mentioned that they need to compare the details of the two graphs carefully to give a final decision in the cases GEGraph fails. 
	Fig. \ref{fig:study 2} presents user ratings of the two graph exploration designs with a 5-point Likert scale, which indicate the high usability of the examples. Most participants showed great interests in using the applications to explore graphs. Several of them praised for the design of integrating node aggregation with Focus+Context interactions, which enables users to explore details within the global content. Some participants commended that they can obtain different insights from the three strategies in related nodes searching.

\begin{figure}[t]
	\setlength{\abovecaptionskip}{2pt}
	\setlength{\belowcaptionskip}{2pt}
	\centering
	\subfigure[Les Mis{\'e}rables dataset]{
		\includegraphics[width=0.23\textwidth]{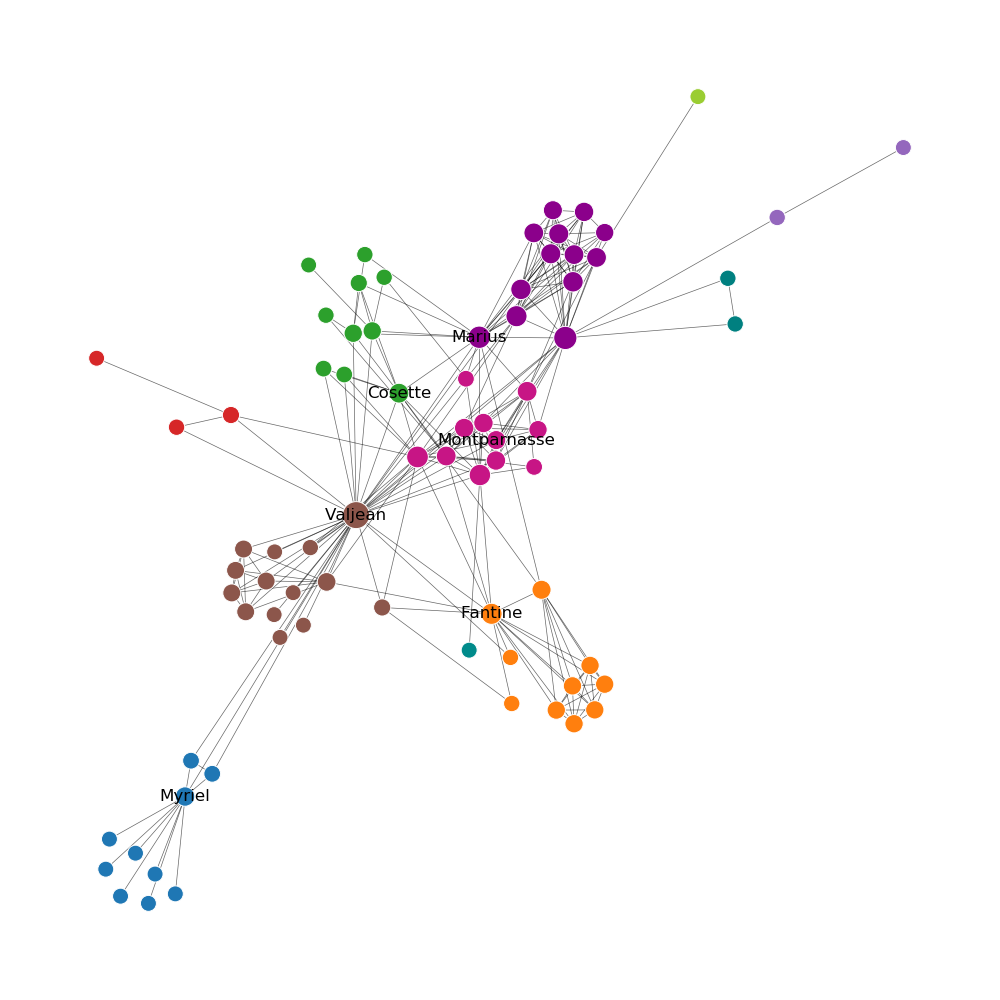}}
	\subfigure[Science dataset]{
		\includegraphics[width=0.23\textwidth]{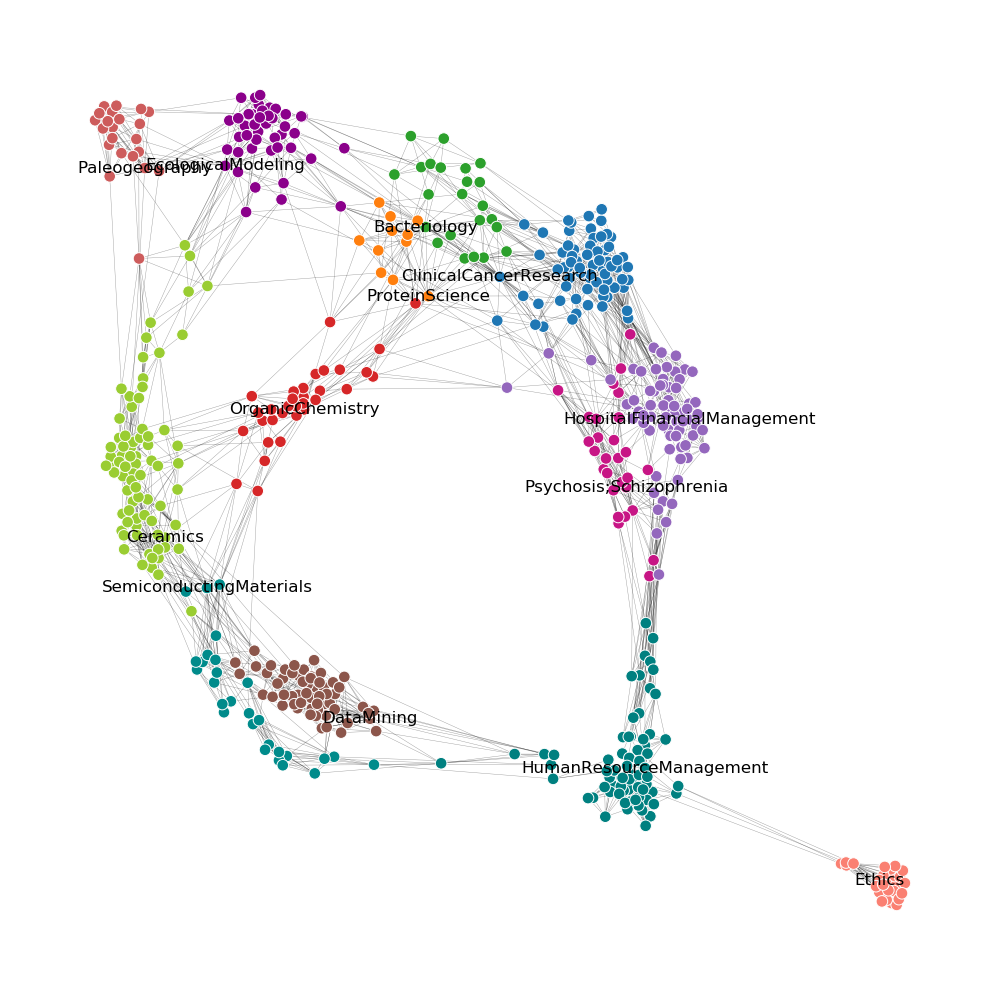}}	
	\caption{
		Use cases. (a) Nodes of major characters in Les Mis{\'e}rables are labeled with character names;
		(b) Representative subdisciplines in major discipline groups are labeled with names.
	}
	\label{fig:case_miserables}
	\vspace{-0.3cm}
\end{figure}

\subsection{Case Studies}\label{sec:case}
\textbf{Les Mis{\'e}rables:}
This is a graph of characters from the novel Les Mis{\'e}rables. An edge in the graph linking two nodes indicates that the two characters appear in the same chapter. Nodes are labeled with groups in that the characters are involved. Fig.~\ref{fig:case_miserables} (a) shows the layout generated by GEGraph, and some nodes are labeled with the character names. Valjean is the central character of this novel, and many other nodes are linked to this node, so it is placed at the center of the layout. Myriel, Cosette, Marius, and Fantine are some other important characters, and they are directly linked to Valjean, but they are in different plots, so they are placed in different communities.
Additionally, Cosette and Marius are married in the story, so they are relatively close in the two communities.

\textbf{Science:}
This graph is a map of subdisciplines of science, where nodes are the disciplines and edges represent that there exist inter-disciplinary works published between them. Each subdiscipline belongs to a major discipline. 
We label nodes with the highest degree in each community. In Fig.~\ref{fig:case_miserables} (b), the overall relations of the subdisciplines are intuitive. In detail, Semiconducting Materials and Data Mining are both about computers, so the disciplines they belong to are connected, and the communities are placed closely.
Protein Science, Clinical Cancer Research, Hospital Financial Management, and Psychosis are all about medical science, and these communities are placed closely in the layout.

\begin{figure}[t]
	\setlength{\abovecaptionskip}{1pt}
	\setlength{\belowcaptionskip}{2pt}
	\hspace{0.9in}
	K-K
	\hspace{0.7in}
	F-R
	\hspace{0.65in}
	SGD
	
	\vspace{0in}
	\centering
	\begin{minipage}[b]{.15\linewidth}
		Original
		\vspace{1.1cm}
	\end{minipage}
	\subfigure{
		\includegraphics[width=0.11\textwidth]{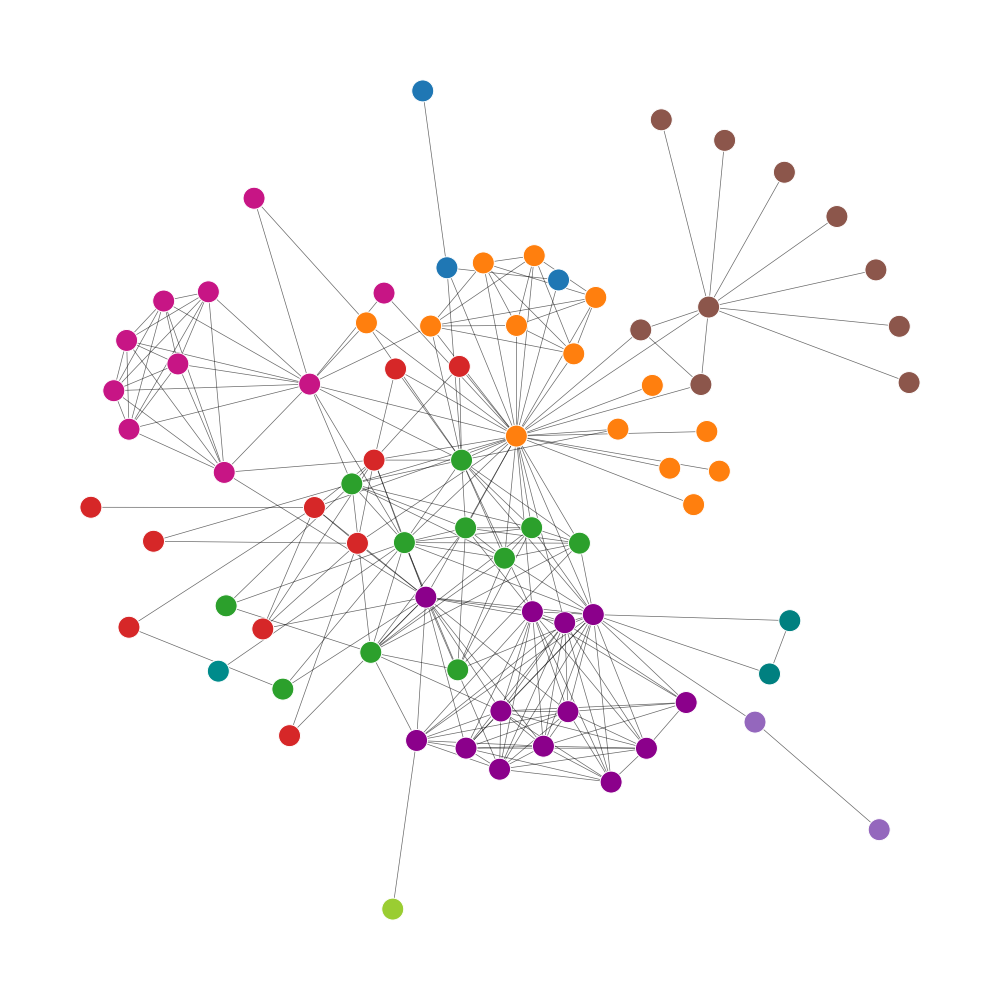}}
	\hspace{0in}
	\subfigure{
		\includegraphics[width=0.11\textwidth]{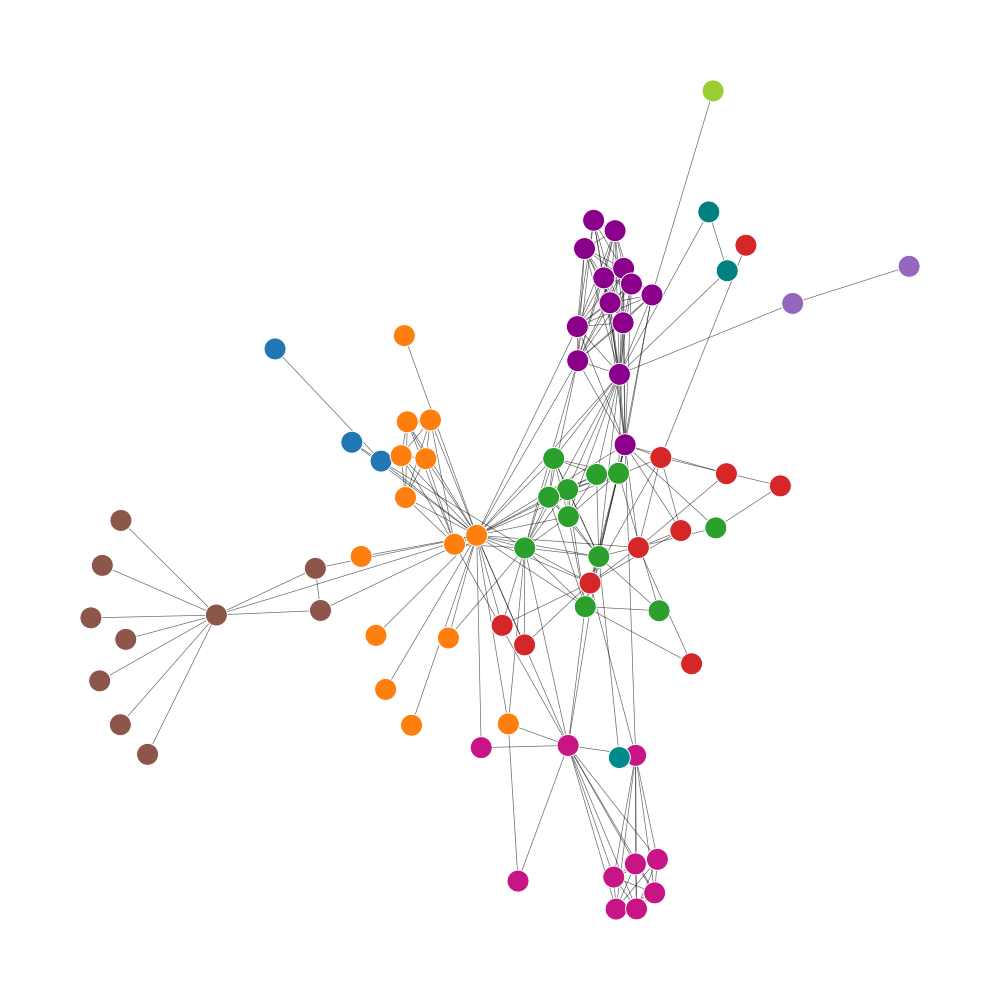}}
	\hspace{0in}
	\subfigure{
		\includegraphics[width=0.11\textwidth]{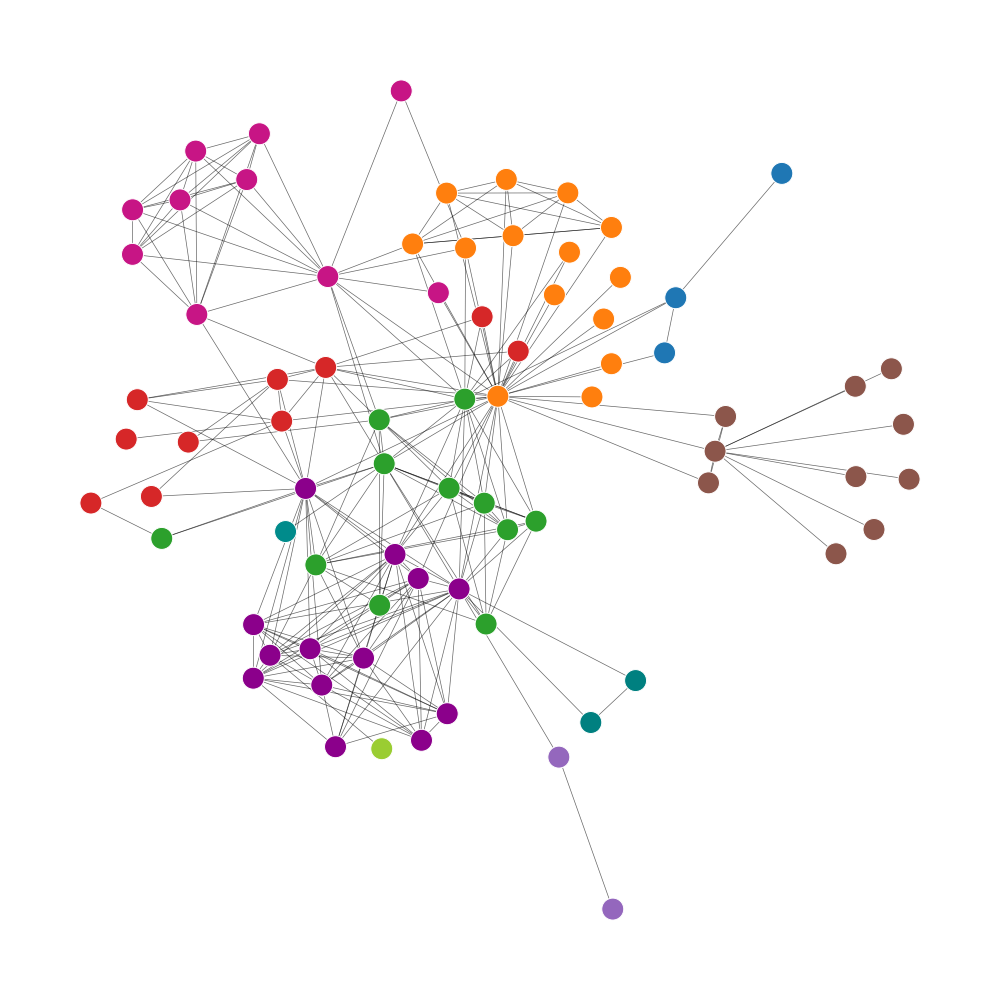}}
	
	\vspace{-0.1in}
	\begin{minipage}[b]{.15\linewidth}
		DeepWalk
		\vspace{1.1cm}
	\end{minipage}
	\subfigure{
		\includegraphics[width=0.11\textwidth]{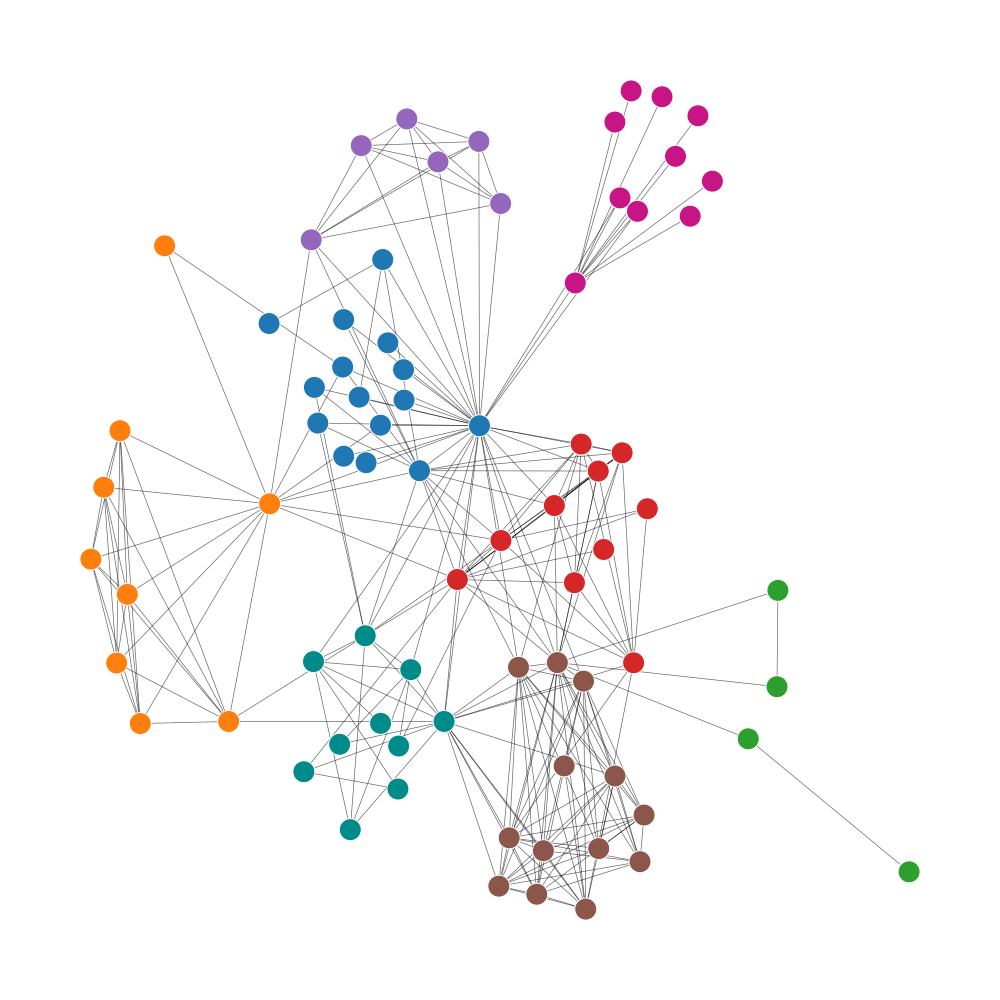}}
	\hspace{0in}
	\subfigure{
		\includegraphics[width=0.11\textwidth]{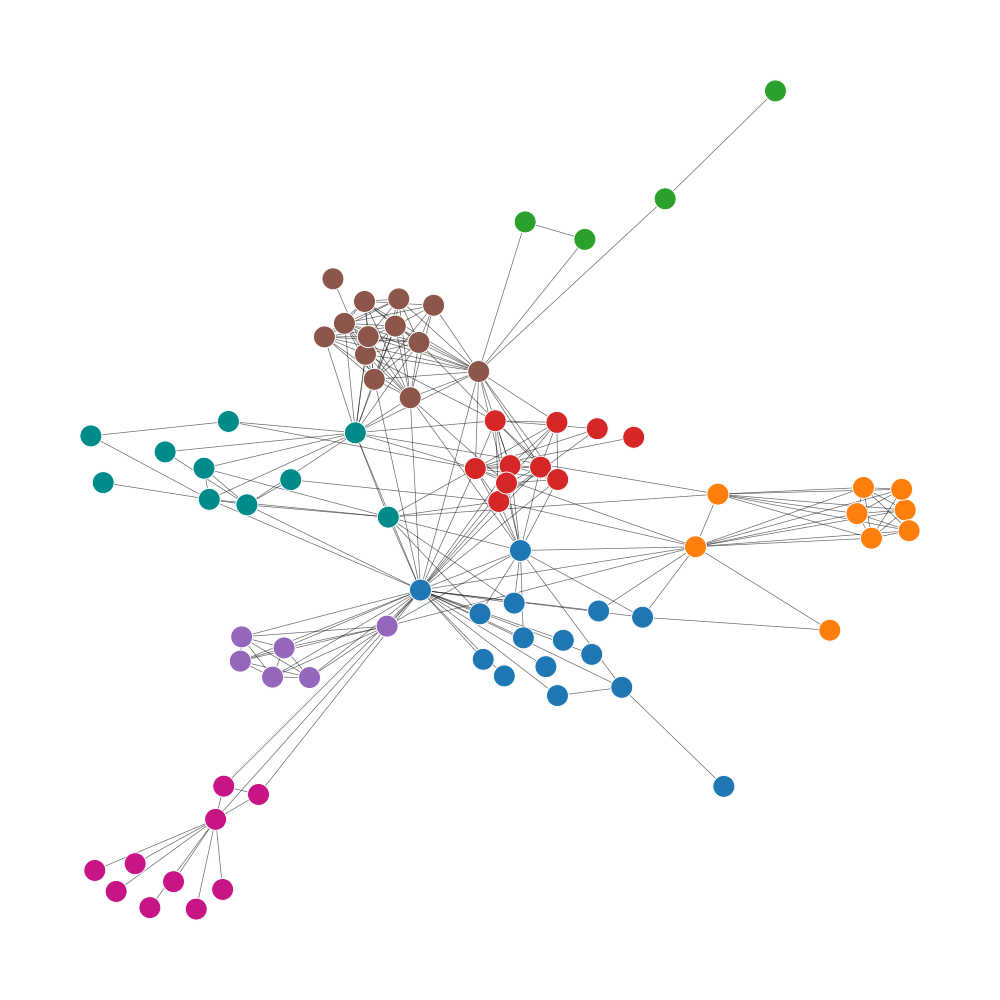}}
	\hspace{0in}
	\subfigure{
		\includegraphics[width=0.11\textwidth]{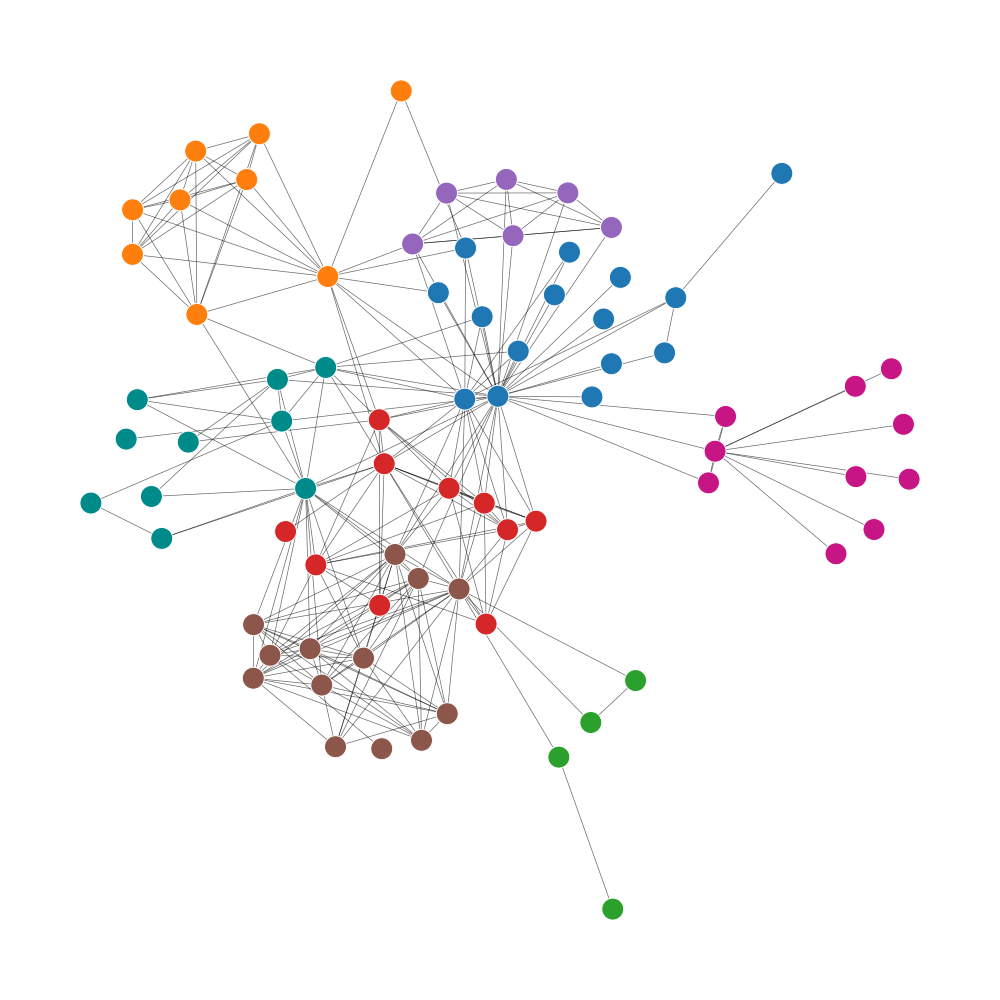}}
	
	\vspace{-0.1in}
	\begin{minipage}[b]{.15\linewidth}
		node2vec
		\vspace{1.1cm}
	\end{minipage}
	\subfigure{
		\includegraphics[width=0.11\textwidth]{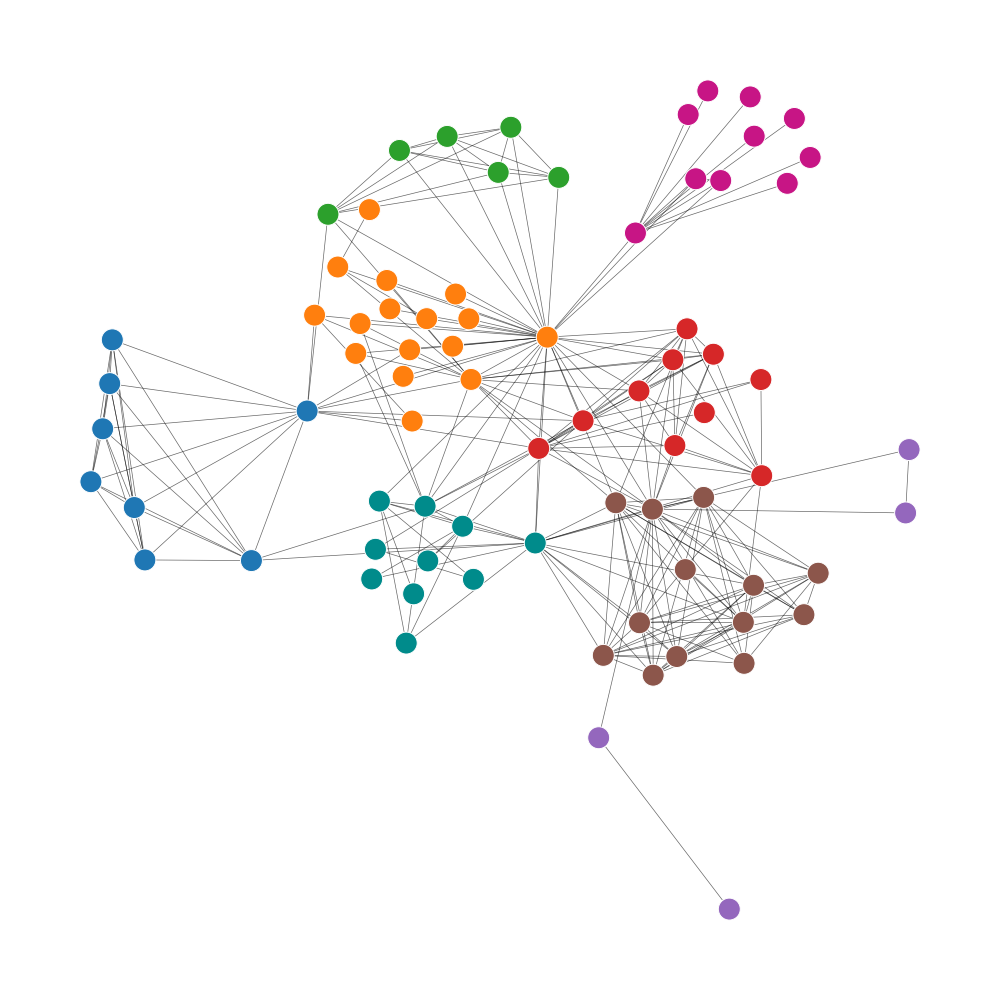}}
	\hspace{0in}
	\subfigure{
		\includegraphics[width=0.11\textwidth]{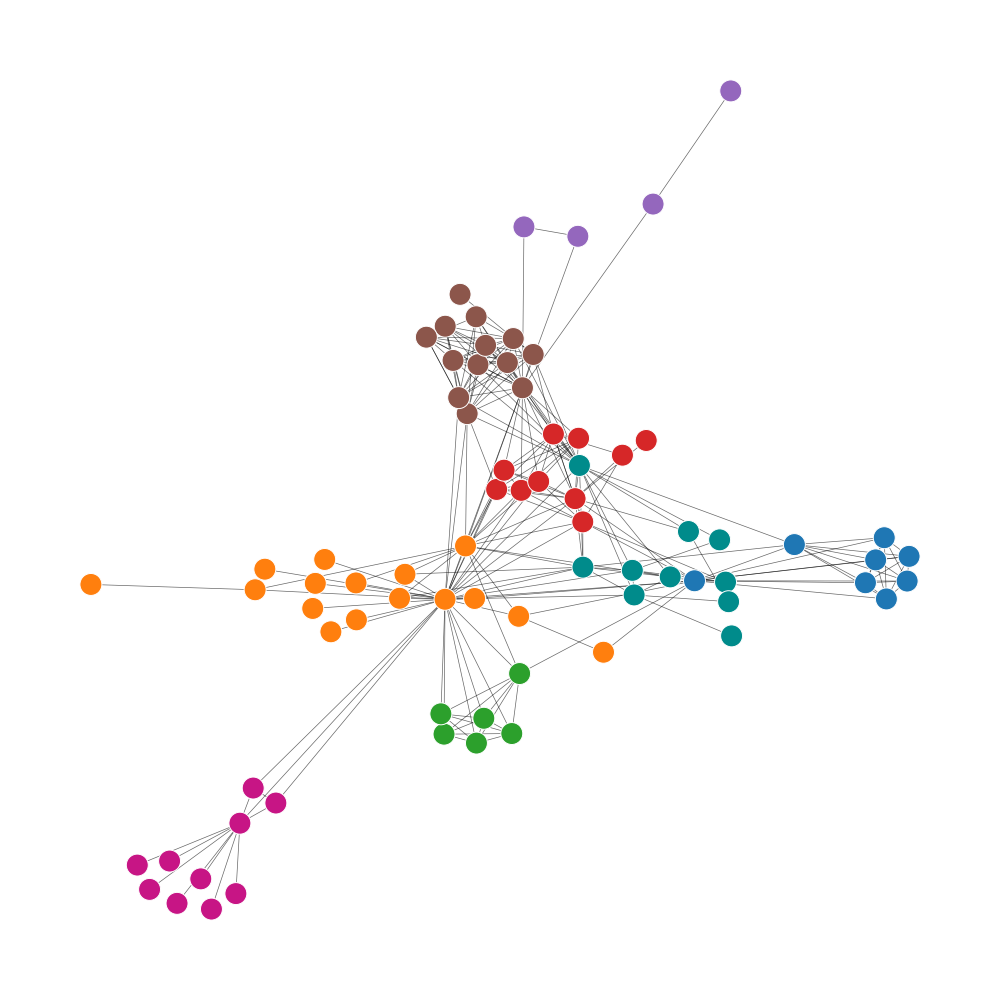}}
	\hspace{0in}
	\subfigure{
		\includegraphics[width=0.11\textwidth]{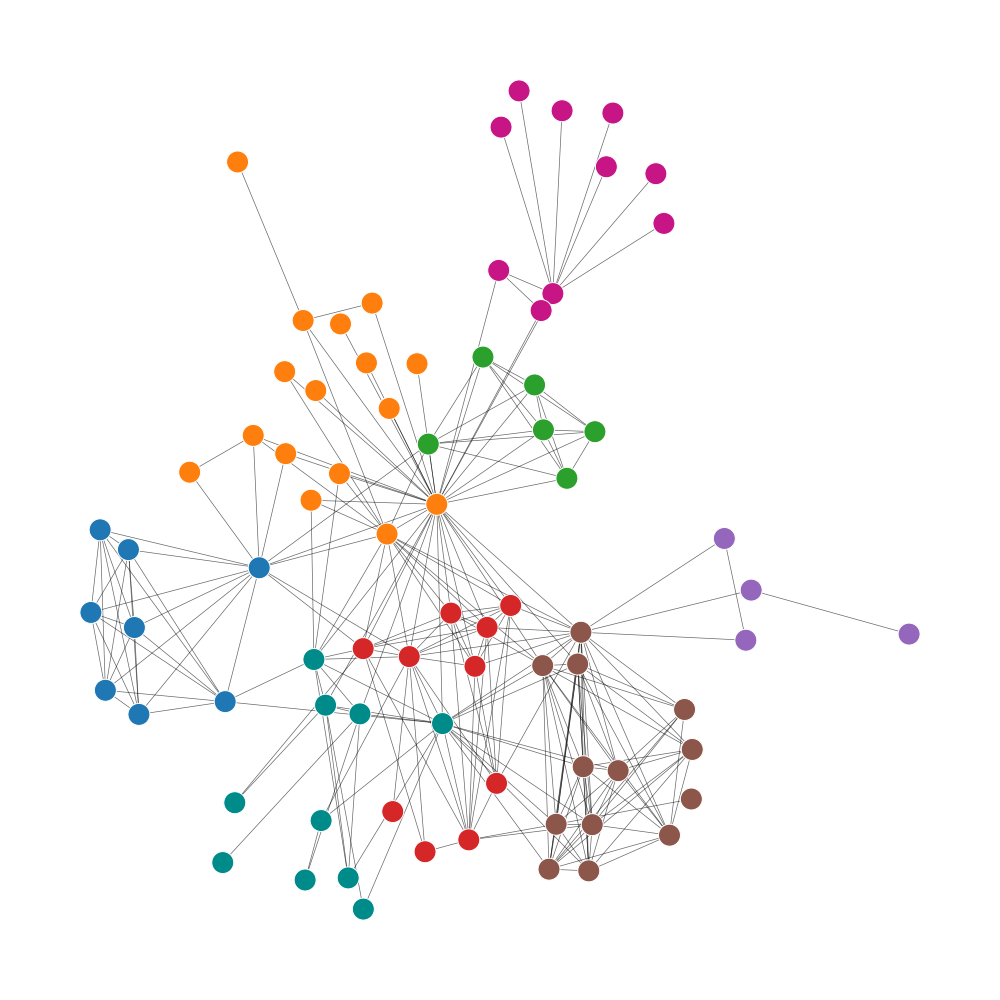}}
	
	\vspace{-0.1in}
	\begin{minipage}[b]{.15\linewidth}
		LINE
		\vspace{1.1cm}
	\end{minipage}
	\subfigure{
		\includegraphics[width=0.11\textwidth]{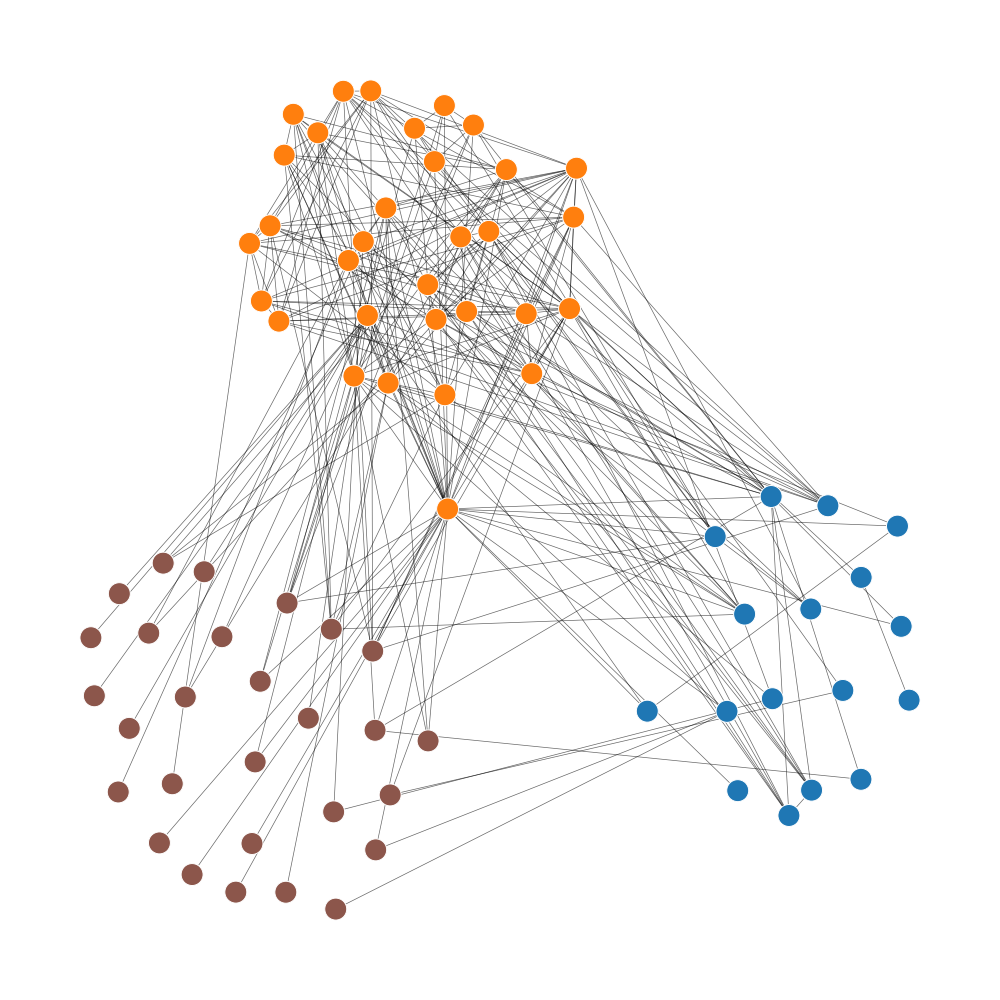}}
	\hspace{0in}
	\subfigure{
		\includegraphics[width=0.11\textwidth]{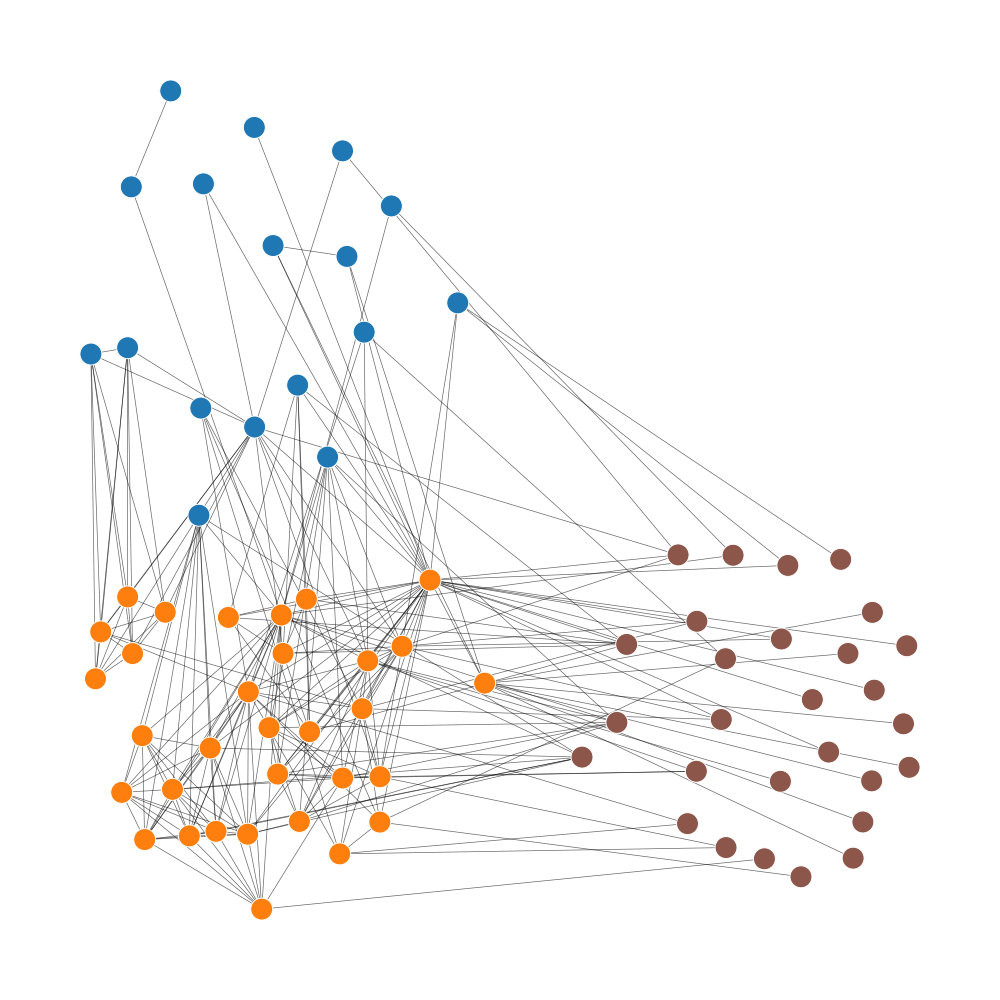}}
	\hspace{0in}
	\subfigure{
		\includegraphics[width=0.11\textwidth]{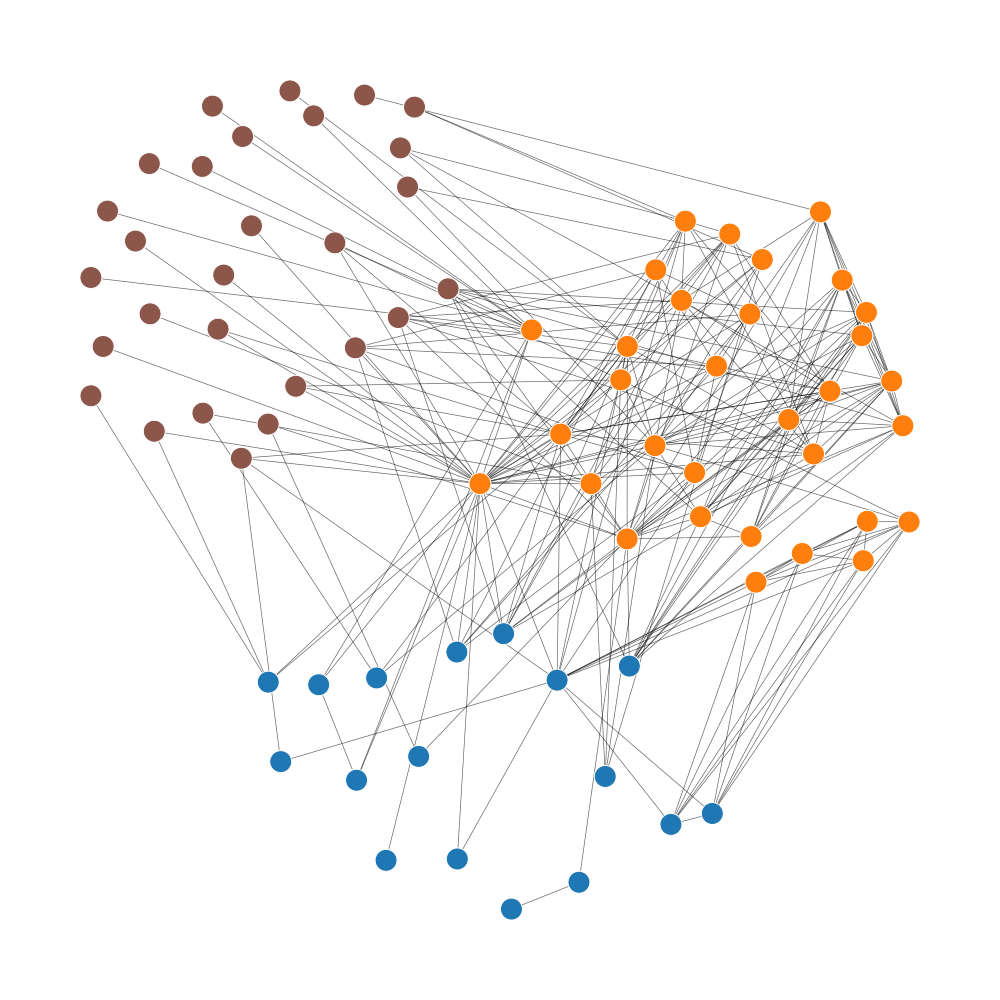}}
	
	\vspace{-0.1in}
	\begin{minipage}[b]{.15\linewidth}
		struc2vec
		\vspace{1.1cm}
	\end{minipage}
	\subfigure{
		\includegraphics[width=0.11\textwidth]{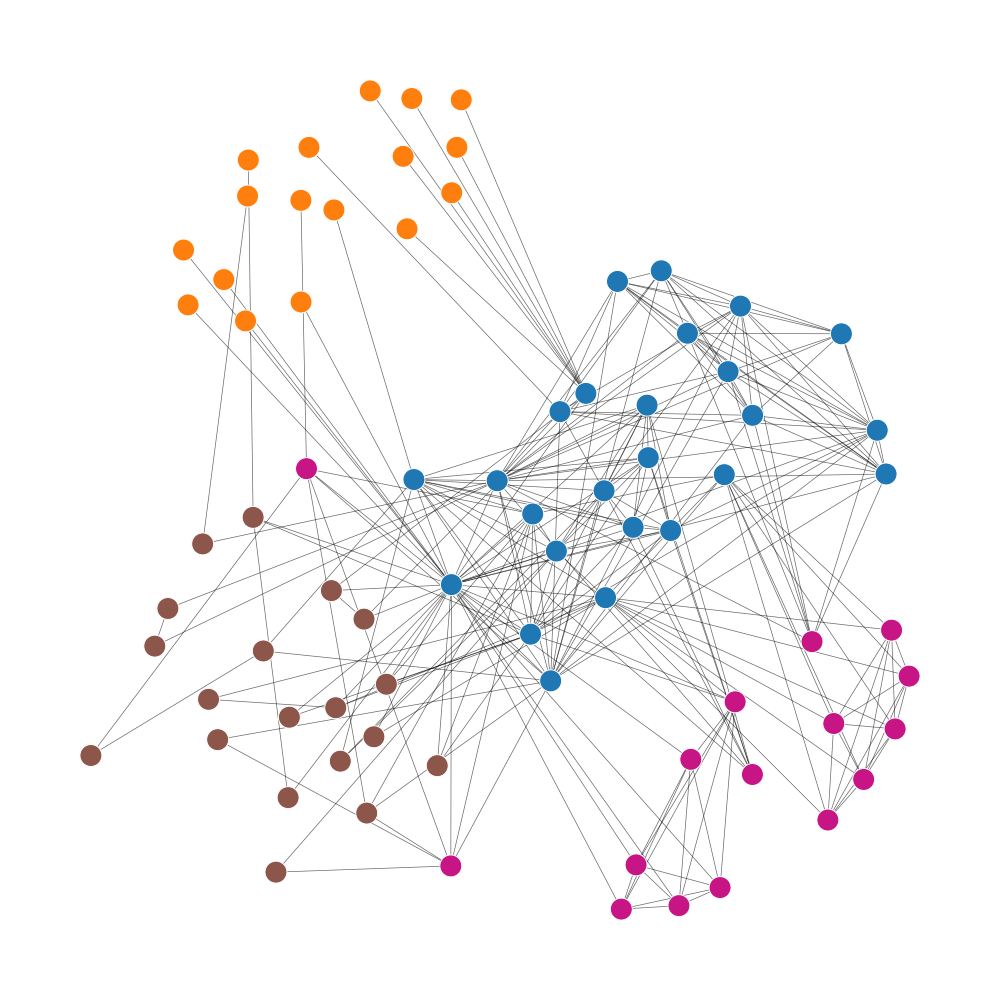}}
	\hspace{0in}
	\subfigure{
		\includegraphics[width=0.11\textwidth]{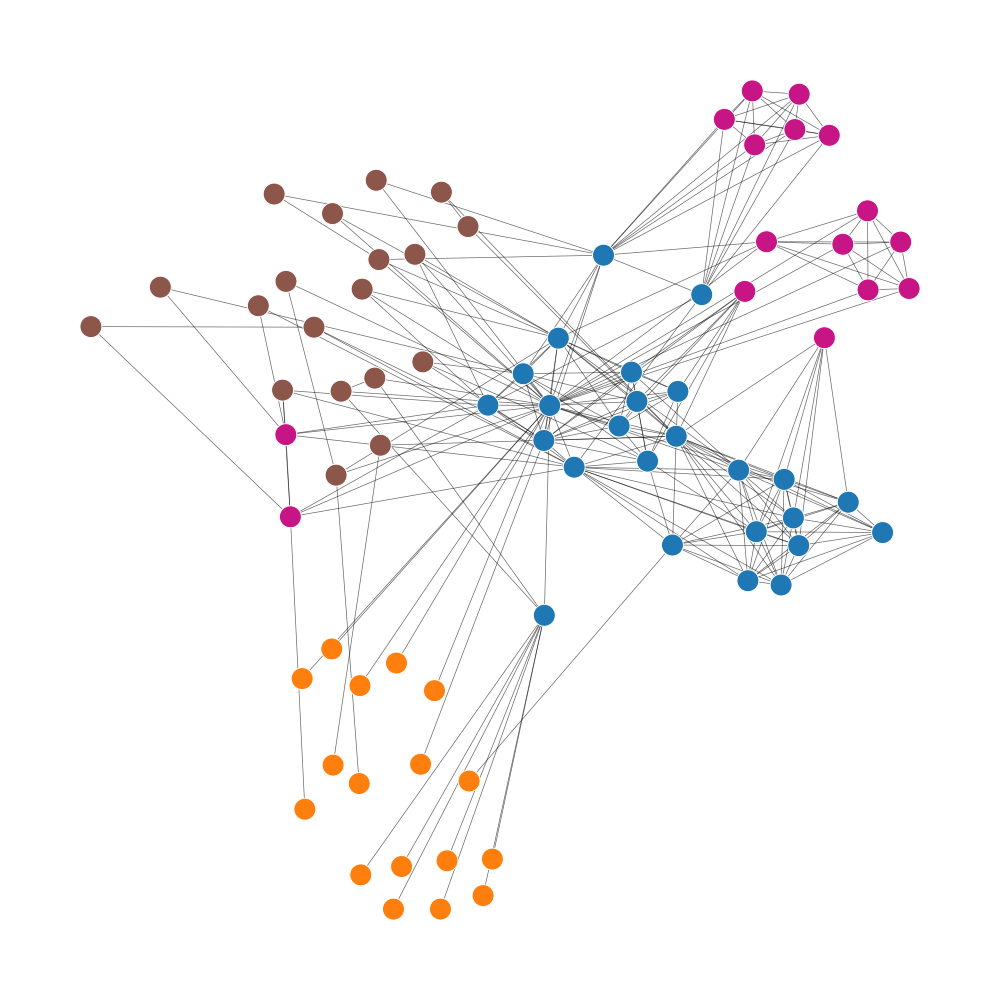}}
	\hspace{0in}
	\subfigure{
		\includegraphics[width=0.11\textwidth]{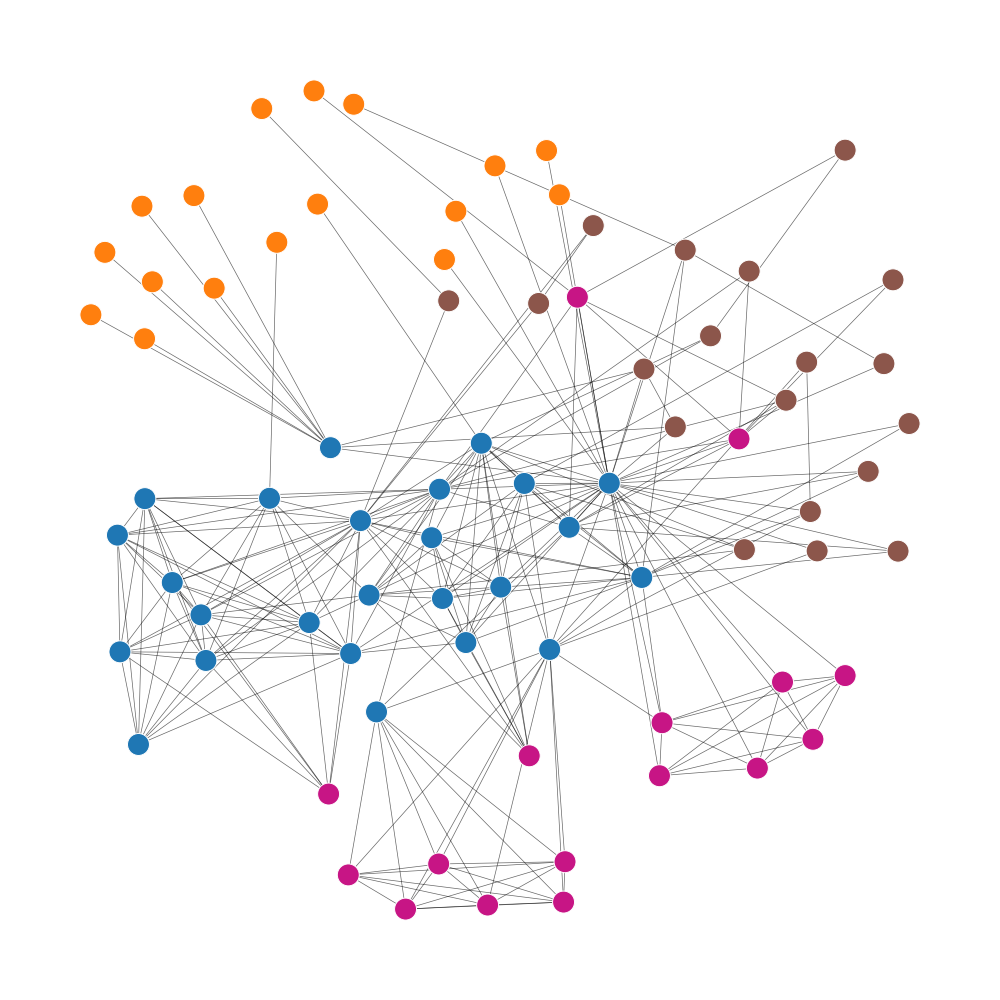}}
	\caption{Combination of different layout and embedding methods in our flexible pipeline with the Les Mis{\'e}rables dataset.
		Embedding methods are DeepWalk, LINE, node2vec and struc2vec,
		and layout methods are K-K, F-R and SGD.
	}
	\label{fig:combination_collection}
	\vspace{-0.2cm}
\end{figure} 

\section{Discussion}\label{sec:discussion}
\noindent
\textbf{Transferability to more embedding methods and layout algorithms. }
According to the flexible design of our pipeline, various graph embedding methods can be leveraged to enhance the graph layout result, and the F-R algorithm can also be replaced. 
Illustrated with examples, as shown in Fig. \ref{fig:combination_collection}, we use DeepWalk~\cite{ozzi2014deepwalk}, LINE~\cite{tang2015line}, node2vec~\cite{grover2016node2vec}, and Struc2vec~\cite{ribeiro2017struc2vec} as embedding methods, and K-K~\cite{kk1989graph}, F-R~\cite{fruchterman1991graph}, and SGD~\cite{graphSGD2018} as layout algorithms.
Though these methods are based on different principles, all of them can process weighted undirected graphs, and thus any of the embedding and layout methods above can be combined arbitrarily.  
Considering the effect and efficiency, we finally chose the combination of node2vec-a and F-R based on our analysis. However, we argue that this combination is not always superior, and we hope future work can find more effective and efficient choices, applying some advanced attribute-first embedding methods~\cite{Zhang2018c,Dong2017,gao2018deep,yang2018binarized}.
It would also be interesting to investigate whether a well-adjusted choice of modules for embedding and layout can improve the results significantly. Additionally, we note that GEGraph does not explicitly strive for minimizing edge crossings, which is mainly due to the use of F-R algorithm~\cite{fruchterman1991graph}. In the future, we can integrate other layout algorithms in the pipeline for certain aesthetic goals (e.g., reduce edge crossings, make edge lengths uniform, etc.).

\begin{figure}[t]
	\setlength{\abovecaptionskip}{-5pt}
	\setlength{\belowcaptionskip}{3pt}
	\centering
	\subfigure[time cost (seconds)]{
		\includegraphics[bb=0 -50 420 380, width=0.22\textwidth]{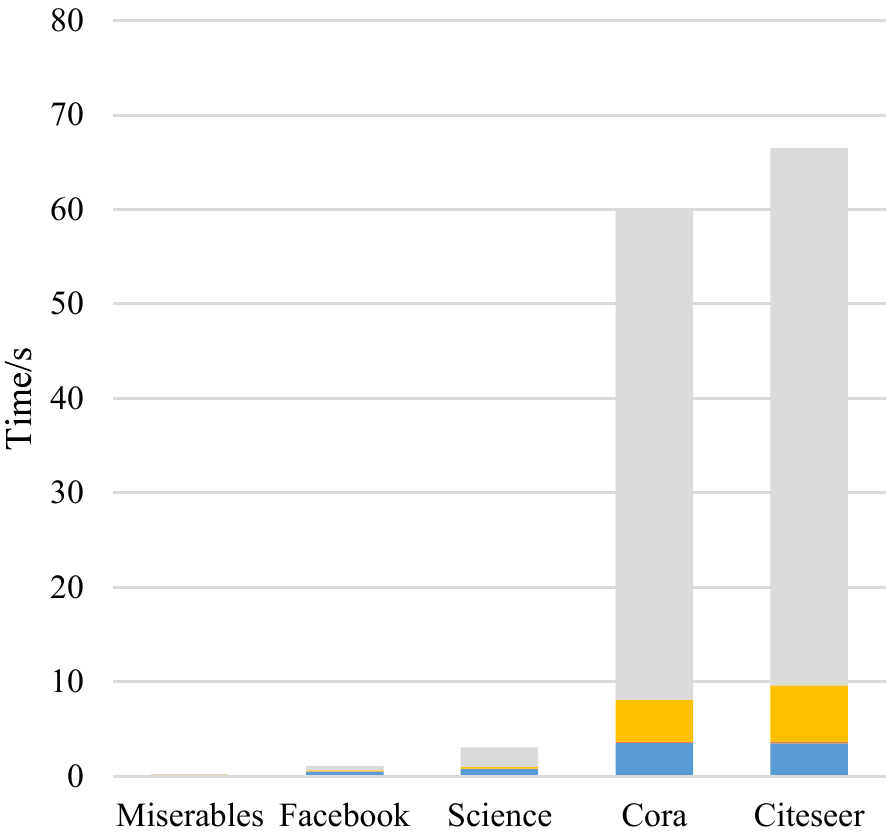}}
	\hspace{0.2in}
	\subfigure[time cost (percentage)]{
		\includegraphics[bb=0 -50 420 380, width=0.22\textwidth]{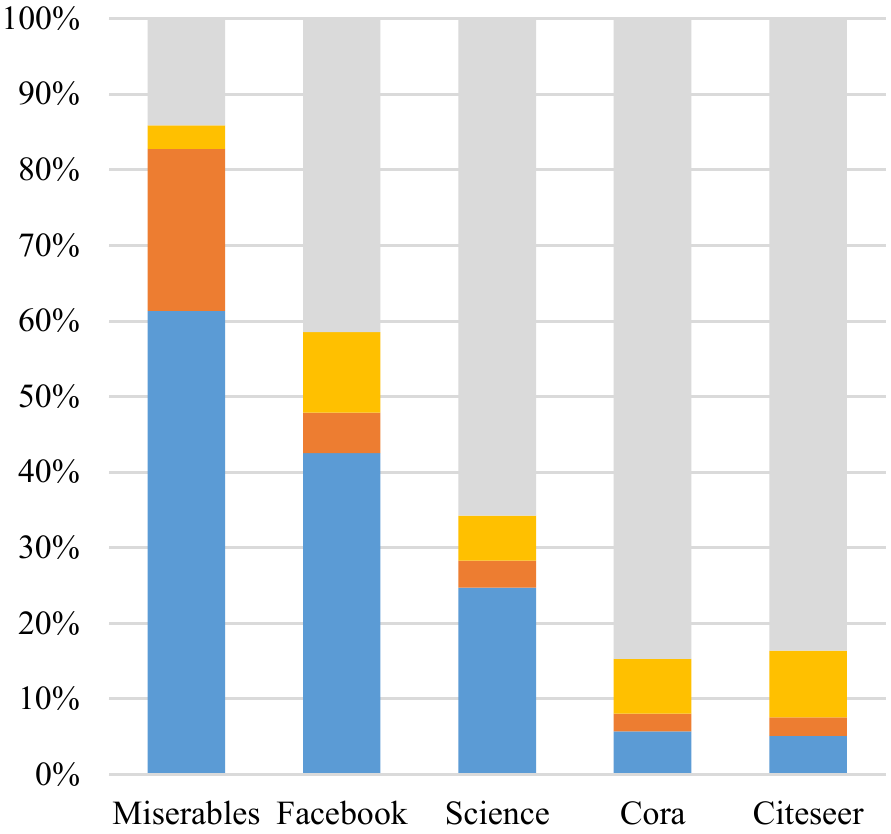}}
	
	\subfigure{
		\includegraphics[bb=-30 -53 490 -73, width=0.48\textwidth]{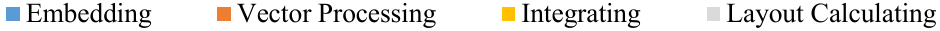}}
	\caption{
		Time cost analysis. 
		(a) displays the seconds of each processing part.
		(b) is the percentage diagram of in (a). Small graphs can be handled rapidly, and the extra time introduced by embedding accounts for a small proportion in large graphs.
	}
	\label{fig:time_costs}
	\vspace{-0.2cm}
\end{figure}

\noindent
\textbf{Scalability and time cost. }
Although graph extension in our pipeline will increase the graph size, the newly introduced virtual nodes are only processed by the embedding method. The layout calculation process handles the graph with the same size as the original one.
Fig.~\ref{fig:time_costs} shows the time cost of our method with different datasets in the computation environment described in Section \ref{sec:setting}. 
In small datasets, the total time cost is small (< 1s), and the proportion of extra time introduced by embedding to the total time cost decreases with the increase of the data scale.
In most situations, the time cost is acceptable for achieving a significantly improved layout result.
Additionally, our embedding method uses the random walk strategy to sample the graph. If switching to matrix factorization-based embedding methods (e.g., HSCA~\cite{zhang2016homophily}) or deep learning-based layout algorithms (e.g., GraphTSNE~\cite{Leow2019}), the time cost will be more considerable. 
Furthermore, optimized embedding methods can be used with small iteration parameters for time-sensitive situations. Users can also adopt a heavy but accurate embedding model to achieve higher practicability~\cite{Zhang2018c,Dong2017}. It is a trade-off between time and quality.


\noindent
\textbf{Degree of automation. }
Our approach does not require users to manually fine-tune the graph layout like PH~\cite{suh2019persistent} and MagnetViz~\cite{Spritzer2012}, nor does it need to adjust complex parameters like DRGraph~\cite{Zhu2021} (including the size of k-order nearest neighbors, the number and weight of negative samples, the iteration number, etc.).
Compared to methods that rely on interactive manipulations~\cite{Spritzer2008,Spritzer2012,Jusufi2013,suh2019persistent}, our approach is self-automatic. Graph embedding, integrating, and layout involve a small set of parameters, as described in Section \ref{sec:setting}.
In general cases, GEGraph can achieve desirable results with a default configuration of parameters, which was tested on a variety of datasets covering different connectivities (connected and non-connected graphs), attribute types (node with single, multiple, or no attributes; nominal and quantitative attributes), data sizes (tens to thousands of nodes or edges), etc.
A major parameter modification is only required when conducting precise adjustments.

\noindent
\textbf{Graph type. }
In our pipeline, all the experiments are conducted on the undirected graph, so the similarity matrix and adjacency matrix can be summed. However, for directed graphs, the adjacency matrix is not symmetric. The calculation of the embedding-enhanced adjacency matrix in our approach will be invalid. So the algorithm should be further reconsidered to support more types of graphs (e.g., directed graphs, weighted graphs, and dynamic graphs). For example, we can modify the adjacency matrix to include edge directions, incorporate weights into the similarity matrix, or introduce a new matrix to encode more information.
Additionally, our discretization of continuous attributes introduces a certain information loss. In the future, we can explore how to better embed continuous attributes for more fine-grained similarity comparisons.

\section{Conclusion}
We propose a flexible embedding-based pipeline to support graph exploration. First, we leverage visualization-oriented graph embedding to encode structure and attribute information into feature vectors.
Second, we present an embedding-driven layout method, which can achieve aesthetic and community-aware visualizations.
Third, we design two example applications for graph exploration, a layout-preserving aggregation approach and a related nodes searching method. 
The quantitative and qualitative evaluation results confirm the effectiveness of our approach.
The pipeline can also be extended to integrate other embedding methods and layout algorithms and develop more interesting exploration applications.
In general, this paper introduces some attempts for embedding-guided graph exploration. We wish that it could inspire new thoughts in the community.

\section*{Acknowledgements}
{
	The authors would like to thank all the reviewers for their valuable suggestions. 
	The work was supported by the National Natural Science Foundation of China (No. 71690231) and Beijing Key Laboratory of Industrial Bigdata System and Application.
}

\bibliographystyle{abbrv}
\bibliography{template}

\begin{thebibliography}{10}

\bibitem{ahmed2018learning}
N.~Ahmed, R.~A. Rossi, J.~Lee, T.~Willke, R.~Zhou, X.~Kong, and H.~Eldardiry.
\newblock {Role-based Graph Embeddings}.
\newblock {\em IEEE Trans. Knowl. Data Eng.}, 34(5):2401--2415, 2022.

\bibitem{Archambault2008a}
D.~Archambault, T.~Munzner, and D.~Auber.
\newblock {GrouseFlocks: Steerable exploration of graph hierarchy space}.
\newblock {\em IEEE Trans. Vis. Comput. Graph.}, 14(4):900--913, 2008.

\bibitem{Bannister2013}
M.~J. Bannister, D.~Eppstein, M.~T. Goodrich, and L.~Trott.
\newblock {Force-directed graph drawing using social gravity and scaling}.
\newblock {\em Graph Draw. Lect. Notes Comput. Sci.}, 7704:414--425, 2013.

\bibitem{10.1109/TVCG.2008.117}
A.~Barsky, T.~Munzner, J.~Gardy, and R.~Kincaid.
\newblock {Cerebral: Visualizing multiple experimental conditions on a graph
  with biological context}.
\newblock {\em IEEE Trans. Vis. Comput. Graph.}, 14(6):1253--1260, 2008.

\bibitem{Bezerianos2010}
A.~Bezerianos, F.~Chevalier, P.~Dragicevic, N.~Elmqvist, and J.~Fekete.
\newblock {GraphDice: A System for Exploring Multivariate Social Networks}.
\newblock {\em Comput. Graph. Forum}, 29(3):863--872, aug 2010.

\bibitem{Bharadwaj2022}
A.~Bharadwaj, D.~Gwizdala, Y.~Kim, K.~Luther, and T.~M. Murali.
\newblock {Flud: A Hybrid Crowd–Algorithm Approach for Visualizing Biological
  Networks}.
\newblock {\em ACM Trans. Comput. Interact.}, 29(1):1--53, 2022.

\bibitem{borner2012design}
K.~B{\"{o}}rner, R.~Klavans, M.~Patek, A.~M. Zoss, J.~R. Biberstine, R.~P.
  Light, V.~Larivi{\`{e}}re, and K.~W. Boyack.
\newblock {Design and update of a classification system: The ucsd map of
  science}.
\newblock {\em PLoS One}, 7(7):1--10, 2012.

\bibitem{Brandes2007}
U.~Brandes and C.~Pich.
\newblock {Eigensolver Methods for Progressive Multidimensional Scaling of
  Large Data}.
\newblock In {\em Graph Draw. Lect. Notes Comput. Sci.}, pages 42--53. Springer
  Berlin Heidelberg, Berlin, Heidelberg, 2007.

\bibitem{Buschel2019}
W.~Buschel, S.~Vogt, and R.~Dachselt.
\newblock {Augmented reality graph visualizations}.
\newblock {\em IEEE Comput. Graph. Appl.}, 39(3):29--40, 2019.

\bibitem{cai2018comprehensive}
H.~Cai, V.~W. Zheng, and K.~C.~C. Chang.
\newblock {A Comprehensive Survey of Graph Embedding: Problems, Techniques, and
  Applications}.
\newblock {\em IEEE Trans. Knowl. Data Eng.}, 30(9):1616--1637, 2018.

\bibitem{PMID:30136986}
W.~Chen, F.~Guo, D.~Han, J.~Pan, X.~Nie, J.~Xia, and X.~Zhang.
\newblock {Structure-Based Suggestive Exploration: A New Approach for Effective
  Exploration of Large Networks}.
\newblock {\em IEEE Trans. Vis. Comput. Graph.}, 25(1):555--565, 2019.

\bibitem{Chen2019g}
Y.~Chen, Z.~Guan, R.~Zhang, X.~Du, and Y.~Wang.
\newblock {A survey on visualization approaches for exploring association
  relationships in graph data}.
\newblock {\em J. Vis.}, 22(3):625--639, 2019.

\bibitem{Cohen1997}
J.~D. Cohen.
\newblock {Drawing graphs to convey proximity}.
\newblock {\em ACM Trans. Comput. Interact.}, 4(3):197--229, sep 1997.

\bibitem{craven1998learning}
M.~Craven, A.~McCallum, D.~PiPasquo, T.~Mitchell, and D.~Freitag.
\newblock Learning to extract symbolic knowledge from the world wide web.
\newblock Technical report, Carnegie-mellon univ pittsburgh pa school of
  computer Science, 1998.

\bibitem{Dong2017}
Y.~Dong, N.~V. Chawla, and A.~Swami.
\newblock {Metapath2vec: Scalable representation learning for heterogeneous
  networks}.
\newblock In {\em KDD'17}, pages 135--144. ACM, 2017.

\bibitem{edgemaps2011}
M.~D{\"{o}}rk, S.~Carpendale, and C.~Williamson.
\newblock {EdgeMaps: visualizing explicit and implicit relations}.
\newblock In {\em VDA'11}, jan 2011.

\bibitem{dunne2015readability}
C.~Dunne, S.~I. Ross, B.~Shneiderman, and M.~Martino.
\newblock {Readability metric feedback for aiding node-link visualization
  designers}.
\newblock {\em IBM J. Res. Dev.}, 59(2-3):11--14, 2015.

\bibitem{Fischer2021}
M.~T. Fischer, A.~Frings, D.~A. Keim, and D.~Seebacher.
\newblock {Towards a Survey on Static and Dynamic Hypergraph Visualizations}.
\newblock In {\em VIS'21}. IEEE, 2021.

\bibitem{fruchterman1991graph}
T.~M. Fruchterman and E.~M. Reingold.
\newblock {Graph drawing by force‐directed placement}.
\newblock {\em Softw. Pract. Exp.}, 21(11):1129--1164, 1991.

\bibitem{Gajer2001}
P.~Gajer, M.~T. Goodrich, and S.~G. Kobourov.
\newblock {A Multi-dimensional Approach to Force-Directed Layouts of Large
  Graphs}.
\newblock In {\em Graph Draw. Lect. Notes Comput. Sci.}, pages 211--221. 2001.

\bibitem{gao2018deep}
H.~Gao and H.~Huang.
\newblock {Deep attributed network embedding}.
\newblock In {\em IJCAI'18}. Morgan Kaufmann, 2018.

\bibitem{Gehlenborg2010}
N.~Gehlenborg, S.~I. O'Donoghue, N.~S. Baliga, A.~Goesmann, M.~A. Hibbs,
  H.~Kitano, O.~Kohlbacher, H.~Neuweger, R.~Schneider, D.~Tenenbaum, and A.~C.
  Gavin.
\newblock {Visualization of omics data for systems biology}.
\newblock {\em Nat. Methods}, 7(3):S56--S68, 2010.

\bibitem{10.1109/TVCG.2013.223}
S.~Ghani, B.~C. Kwon, S.~Lee, J.~S. Yi, and N.~Elmqvist.
\newblock {Visual analytics for multimodal social network analysis: A design
  study with social scixentists}.
\newblock {\em IEEE Trans. Vis. Comput. Graph.}, 19(12):2032--2041, 2013.

\bibitem{gibson2013survey}
H.~Gibson, J.~Faith, and P.~Vickers.
\newblock {A survey of two-dimensional graph layout techniques for information
  visualisation}.
\newblock {\em Inf. Vis.}, 12(3-4):324--357, 2013.

\bibitem{Gibson2016}
H.~Gibson and P.~Vickers.
\newblock {Using adjacency matrices to lay out larger small-world networks}.
\newblock {\em Appl. Soft Comput. J.}, 42:80--92, 2016.

\bibitem{Gibson2017}
H.~Gibson and P.~Vickers.
\newblock {graphTPP: A multivariate based method for interactive graph layout
  and analysis}.
\newblock {\em arXiv}, 2017.

\bibitem{grover2016node2vec}
A.~Grover and J.~Leskovec.
\newblock {Node2vec: Scalable feature learning for networks}.
\newblock In {\em KDD'16}. ACM, 2016.

\bibitem{PMID:19834170}
D.~Guo.
\newblock {Flow mapping and multivariate visualization of large spatial
  interaction data}.
\newblock {\em IEEE Trans. Vis. Comput. Graph.}, 15(6):1041--1048, 2009.

\bibitem{Hachul2005}
S.~Hachul and M.~J{\"{u}}nger.
\newblock {Drawing Large Graphs with a Potential-Field-Based Multilevel
  Algorithm}.
\newblock In {\em Graph Draw. Lect. Notes Comput. Sci.}, pages 285--295. 2005.

\bibitem{haleem2019evaluating}
H.~Haleem, Y.~Wang, A.~Puri, S.~Wadhwa, and H.~Qu.
\newblock {Evaluating the Readability of Force Directed Graph Layouts: A Deep
  Learning Approach}.
\newblock {\em IEEE Comput. Graph. Appl.}, 39(4):40--53, 2019.

\bibitem{Harel2002}
D.~Harel and Y.~Koren.
\newblock {Graph Drawing by High-Dimensional Embedding}.
\newblock {\em J. Graph Algorithms Appl.}, 8(2):195--214, jun 2004.

\bibitem{Henderson2012}
K.~Henderson, B.~Gallagher, T.~Eliassi-Rad, H.~Tong, S.~Basu, L.~Akoglu,
  D.~Koutra, C.~Faloutsos, and L.~Li.
\newblock {RolX: Structural role extraction {\&} mining in large graphs}.
\newblock In {\em KDD'12}, pages 1231--1239. ACM, 2012.

\bibitem{herman2000graph}
I.~Herman, G.~Melan{\c{c}}on, and M.~S. Marshall.
\newblock {Graph visualization and navigation in information visualization: a
  survey}.
\newblock {\em IEEE Trans. Vis. Comput. Graph.}, 6(1):24--43, 2000.

\bibitem{Horak2020}
T.~Horak, P.~Berger, H.~Schumann, R.~Dachselt, and C.~Tominski.
\newblock {Responsive Matrix Cells: A Focus+Context Approach for Exploring and
  Editing Multivariate Graphs}.
\newblock {\em IEEE Trans. Vis. Comput. Graph.}, 27(2):1644--1654, 2021.

\bibitem{Itoh2015}
T.~Itoh and K.~Klein.
\newblock {Key-node-separated graph clustering and layouts for human
  relationship graph visualization}.
\newblock {\em IEEE Comput. Graph. Appl.}, 35(6):30--40, 2015.

\bibitem{Jusufi2013}
I.~Jusufi, A.~Kerren, and B.~Zimmer.
\newblock {Multivariate network exploration with JauntyNets}.
\newblock In {\em IV'13}, pages 19--27, 2013.

\bibitem{kk1989graph}
T.~Kamada and S.~Kawai.
\newblock {An algorithm for drawing general undirected graphs}.
\newblock {\em Inf. Process. Lett.}, 31(1):7--15, 1989.

\bibitem{Kerren2014}
A.~Kerren, H.~C. Purchase, and M.~O. Ward, editors.
\newblock {\em {Multivariate Network Visualization}}, volume 8380 of {\em
  Lecture Notes in Computer Science}.
\newblock Springer International Publishing, 2014.

\bibitem{knuth1993stanford}
D.~E. Knuth.
\newblock {\em {Stanford GraphBase: A platform for combinatorial algorithms}}.
\newblock ACM-SIAM, 1993.

\bibitem{Kruiger2017}
J.~F. Kruiger, P.~E. Rauber, R.~M. Martins, A.~Kerren, S.~Kobourov, and A.~C.
  Telea.
\newblock {Graph Layouts by t-SNE}.
\newblock {\em Comput. Graph. Forum}, 36(3):283--294, 2017.

\bibitem{PMID:19541911}
M.~Krzywinski, J.~Schein, I.~Birol, J.~Connors, R.~Gascoyne, D.~Horsman, S.~J.
  Jones, and M.~A. Marra.
\newblock {Circos: An information aesthetic for comparative genomics}.
\newblock {\em Genome Res.}, 19(9):1639--1645, 2009.

\bibitem{Kwon2018}
O.~H. Kwon, T.~Crnovrsanin, and K.~L. Ma.
\newblock {What Would a Graph Look Like in this Layout? A Machine Learning
  Approach to Large Graph Visualization}.
\newblock {\em IEEE Trans. Vis. Comput. Graph.}, 24(1):478--488, 2018.

\bibitem{2019arXiv190412225K}
O.~H. Kwon and K.~L. Ma.
\newblock {A Deep Generative Model for Graph Layout}.
\newblock {\em IEEE Trans. Vis. Comput. Graph.}, 26(1):665--675, 2020.

\bibitem{lee2006}
B.~Lee, C.~Plaisant, C.~S. Parr, J.~D. Fekete, and N.~Henry.
\newblock {Task taxonomy for graph visualization}.
\newblock In {\em Proc. BELIV'06}. ACM, 2006.

\bibitem{Lehtinen2012}
P.~Lehtinen, M.~Saarela, and T.~Elomaa.
\newblock {Online ChiMerge Algorithm}.
\newblock In {\em Intell. Syst. Ref. Libr.}, pages 199--216. 2012.

\bibitem{Leow2019}
Y.~Y. Leow, T.~Laurent, and X.~Bresson.
\newblock {GraphTSNE: A Visualization Technique for Graph-Structured Data}.
\newblock In {\em ICLR'19}, 2019.

\bibitem{liao2018attributed}
L.~Liao, X.~He, H.~Zhang, and T.~S. Chua.
\newblock {Attributed Social Network Embedding}.
\newblock {\em IEEE Trans. Knowl. Data Eng.}, 30(12):2257--2270, 2018.

\bibitem{lu2003link}
Q.~Lu and L.~Getoor.
\newblock {Link-based classification}.
\newblock In {\em ICML'03}, 2003.

\bibitem{Martins2017}
R.~M. Martins, J.~F. Kruiger, R.~Minghim, A.~C. Telea, and A.~Kerren.
\newblock {MVN-Reduce : Dimensionality Reduction for the Visual Analysis of
  Multivariate Networks}.
\newblock In {\em EuroVis'17 Short Paper}, pages 10--14, 2017.

\bibitem{leskovec2012learning}
J.~McAuley and J.~Leskovec.
\newblock {Learning to discover social circles in ego networks}.
\newblock In {\em NIPS'12}, pages 539--547, 2012.

\bibitem{Meyer2009MizBee}
M.~Meyer, T.~Munzner, and H.~Pfister.
\newblock {MizBee: A multiscale synteny browsers}.
\newblock {\em IEEE Trans. Vis. Comput. Graph.}, 15(6):897--904, 2009.

\bibitem{Neuweger2009}
H.~Neuweger, M.~Persicke, S.~P. Albaum, T.~Bekel, M.~Dondrup, A.~T.
  H{\"{u}}ser, J.~Winnebald, J.~Schneider, J.~Kalinowski, and A.~Goesmann.
\newblock {Visualizing post genomics data-sets on customized pathway maps by
  ProMeTra – aeration-dependent gene expression and metabolism of
  Corynebacterium glutamicum as an example}.
\newblock {\em BMC Syst. Biol.}, 3(1):82, dec 2009.

\bibitem{noack2003energy}
A.~Noack.
\newblock {An energy model for visual graph clustering}.
\newblock {\em Graph Draw. Lect. Notes Comput. Sci.}, 2912:425--436, 2003.

\bibitem{Noack2005}
A.~Noack and C.~Lewerentz.
\newblock {A space of layout styles for hierarchical graph models of software
  systems}.
\newblock {\em SoftVis'05}, 1(212):155--164, 2005.

\bibitem{nobre2019state}
C.~Nobre, M.~Meyer, M.~Streit, and A.~Lex.
\newblock {The state of the art in visualizing multivariate networks}.
\newblock In {\em Comput. Graph. Forum}, volume~38, pages 807--832. Wiley,
  2019.

\bibitem{Partl2014ConTour}
C.~Partl, A.~Lex, M.~Streit, H.~Strobelt, A.~M. Wassermann, H.~Pfister, and
  D.~Schmalstieg.
\newblock {ConTour: Data-driven exploration of multi-relational datasets for
  drug discovery}.
\newblock {\em IEEE Trans. Vis. Comput. Graph.}, 20(12):1883--1892, 2014.

\bibitem{ozzi2014deepwalk}
B.~Perozzi, R.~Al-Rfou, and S.~Skiena.
\newblock {DeepWalk: Online learning of social representations}.
\newblock In {\em KDD'14}. ACM, 2014.

\bibitem{Pretorius2008Visual}
A.~J. Pretorius and J.~J. {Van Wijk}.
\newblock {Visual inspection of multivariate graphs}.
\newblock {\em Comput. Graph. Forum}, 27(3):967--974, 2008.

\bibitem{ribeiro2017struc2vec}
L.~F. Ribeiro, P.~H. Saverese, and D.~R. Figueiredo.
\newblock {Struc2vec: Learning node representations from structural identity}.
\newblock In {\em KDD'17}. ACM, 2017.

\bibitem{Robertson2004}
S.~Robertson.
\newblock {Understanding inverse document frequency: on theoretical arguments
  for IDF}.
\newblock {\em J. Doc.}, 60(5):503--520, oct 2004.

\bibitem{Shen2021a}
L.~Shen, E.~Shen, Y.~Luo, X.~Yang, X.~Hu, X.~Zhang, Z.~Tai, and J.~Wang.
\newblock {Towards Natural Language Interfaces for Data Visualization: A
  Survey}.
\newblock {\em IEEE Trans. Vis. Comput. Graph.}, pages 1--20, sep 2022.

\bibitem{Shi2014}
L.~Shi, Q.~Liao, H.~Tong, Y.~Hu, Y.~Zhao, and C.~Lin.
\newblock {Hierarchical focus+context heterogeneous network visualization}.
\newblock In {\em PacificVis'14}, pages 89--96. IEEE, 2014.

\bibitem{SS2006BA}
B.~Shneiderman and A.~Aris.
\newblock {Network visualization by semantic substrates}.
\newblock {\em IEEE Trans. Vis. Comput. Graph.}, 12(5):733--740, 2006.

\bibitem{10.5555/2384008.2384030}
S.~Spanurattana and T.~Murata.
\newblock {Visual analysis of bipartite networks}.
\newblock In {\em ICDM'11}. IEEE, 2011.

\bibitem{Spritzer2008}
A.~S. Spritzer and C.~M. Freitas.
\newblock {A Physics-based Approach for Interactive Manipulation of Graph
  Visualizations}.
\newblock In {\em Proc. AVI'08}, pages 271--278, 2008.

\bibitem{Spritzer2012}
A.~S. Spritzer and C.~M. D.~S. Freitas.
\newblock {Design and evaluation of MagnetViz - A graph visualization tool}.
\newblock {\em IEEE Trans. Vis. Comput. Graph.}, 18(5):822--835, 2012.

\bibitem{Srinivasan2018}
A.~Srinivasan and J.~Stasko.
\newblock {Orko: Facilitating Multimodal Interaction for Visual Exploration and
  Analysis of Networks}.
\newblock {\em IEEE Trans. Vis. Comput. Graph.}, 24(1):511--521, 2018.

\bibitem{suh2019persistent}
A.~Suh, M.~Hajij, B.~Wang, C.~Scheidegger, and P.~Rosen.
\newblock {Persistent Homology Guided Force-Directed Graph Layouts}.
\newblock {\em IEEE Trans. Vis. Comput. Graph.}, 26(1):697--707, 2020.

\bibitem{tang2015line}
J.~Tang, M.~Qu, M.~Wang, M.~Zhang, J.~Yan, and Q.~Mei.
\newblock {LINE: Large-scale information network embedding}.
\newblock In {\em WWW'15}. ACM, 2015.

\bibitem{PMID:26356945}
S.~{Van Den Elzen} and J.~J. {Van Wijk}.
\newblock {Multivariate network exploration and presentation: From detail to
  overview via selections and aggregations}.
\newblock {\em IEEE Trans. Vis. Comput. Graph.}, 20(12):2310--2319, 2014.

\bibitem{van2014multivariate}
S.~{Van Den Elzen} and J.~J. {Van Wijk}.
\newblock {Multivariate network exploration and presentation: From detail to
  overview via selections and aggregations}.
\newblock {\em IEEE Trans. Vis. Comput. Graph.}, 20(12):2310--2319, 2014.

\bibitem{vandermaaten08a}
L.~{Van Der Maaten} and G.~Hinton.
\newblock {Visualizing data using t-SNE}.
\newblock {\em J. Mach. Learn. Res.}, 9(86):2579--2625, 2008.

\bibitem{VanHam2006}
F.~van Ham, N.~Krishnan, and Others.
\newblock {Ask-graphview: A large scale graph visualization system}.
\newblock {\em IEEE Trans. Vis. Comput. Graph.}, 12(5):669--676, 2006.

\bibitem{VonLandesberger2011}
T.~von Landesberger, A.~Kuijper, T.~Schreck, J.~Kohlhammer, J.~J. van Wijk,
  J.~D. Fekete, and D.~W. Fellner.
\newblock {Visual analysis of large graphs: State-of-the-art and future
  research challenges}.
\newblock {\em Eurographics Symp. Geom. Process.}, 30(6):1719--1749, 2011.

\bibitem{Walshaw2001}
C.~Walshaw.
\newblock {A Multilevel Algorithm for Force-Directed Graph Drawing}.
\newblock {\em J. Graph Algorithms Appl.}, 7(3):171--182, 2001.

\bibitem{wang2019deepdrawing}
Y.~Wang, Z.~Jin, Q.~Wang, W.~Cui, T.~Ma, and H.~Qu.
\newblock {DeepDrawing: A Deep Learning Approach to Graph Drawing}.
\newblock {\em IEEE Trans. Vis. Comput. Graph.}, 26(1):676--686, 2020.

\bibitem{Wang2016d}
Y.~Wang, Q.~Shen, D.~Archambault, Z.~Zhou, M.~Zhu, S.~Yang, and H.~Qu.
\newblock {AmbiguityVis: Visualization of Ambiguity in Graph Layouts}.
\newblock {\em IEEE Trans. Vis. Comput. Graph.}, 22(1):359--368, 2016.

\bibitem{wang2019interactive}
Y.~Wang, M.~Xue, Y.~Wang, X.~Yan, B.~Chen, C.~W. Fu, and C.~Hurter.
\newblock {Interactive Structure-aware Blending of Diverse Edge Bundling
  Visualizations}.
\newblock {\em IEEE Trans. Vis. Comput. Graph.}, 26(1):687--696, 2020.

\bibitem{wattenberg2006visual}
M.~Wattenberg.
\newblock {Visual exploration of multivariate graphs}.
\newblock In {\em CHI'06}. ACM, 2006.

\bibitem{wongsuphasawat2017visualizing}
K.~Wongsuphasawat, D.~Smilkov, J.~Wexler, J.~Wilson, D.~Man{\'{e}}, D.~Fritz,
  D.~Krishnan, F.~B. Vi{\'{e}}gas, and M.~Wattenberg.
\newblock {Visualizing Dataflow Graphs of Deep Learning Models in TensorFlow}.
\newblock {\em IEEE Trans. Vis. Comput. Graph.}, 24(1):1--12, 2018.

\bibitem{Wu2016}
Y.~Wu, N.~Pitipornvivat, J.~Zhao, S.~Yang, G.~Huang, and H.~Qu.
\newblock {egoSlider: Visual Analysis of Egocentric Network Evolution}.
\newblock {\em IEEE Trans. Vis. Comput. Graph.}, 22(1):260--269, 2016.

\bibitem{Wu2006}
Y.~Wu and M.~Takatsuka.
\newblock {Visualizing multivariate network on the surface of a sphere}.
\newblock In {\em APVis'06}, pages 77--83, 2006.

\bibitem{Wu2008}
Y.~Wu and M.~Takatsuka.
\newblock {Visualizing Multivariate Networks: A Hybrid Approach}.
\newblock In {\em PacificVis'08}, pages 223--230. IEEE, 2008.

\bibitem{Xue2022}
M.~Xue, Y.~Wang, C.~Han, J.~Zhang, Z.~Wang, K.~Zhang, C.~Hurter, J.~Zhao, and
  O.~Deussen.
\newblock {Target Netgrams: An Annulus-Constrained Stress Model for Radial
  Graph Visualization}.
\newblock {\em IEEE Trans. Vis. Comput. Graph.}, pages 1--13, 2022.

\bibitem{yang2015network}
C.~Yang, Z.~Liu, D.~Zhao, M.~Sun, and E.~Y. Chang.
\newblock {Network representation learning with rich text information}.
\newblock In {\em IJCAI'15}. Morgan Kaufmann, 2015.

\bibitem{yang2018binarized}
H.~Yang, S.~Pan, P.~Zhang, L.~Chen, D.~Lian, and C.~Zhang.
\newblock {Binarized attributed network embedding}.
\newblock In {\em ICDM'18}. IEEE, 2018.

\bibitem{Yang2014f}
J.~Yang and J.~Leskovec.
\newblock {Overlapping Communities Explain Core–Periphery Organization of
  Networks}.
\newblock {\em Proc. IEEE}, 102(12):1892--1902, dec 2014.

\bibitem{zhang2016homophily}
D.~Zhang, J.~Yin, X.~Zhu, and C.~Zhang.
\newblock {Homophily, structure, and content augmented network representation
  learning}.
\newblock In {\em ICDM'17}. IEEE, 2017.

\bibitem{Zhang2018c}
D.~Zhang, J.~Yin, X.~Zhu, and C.~Zhang.
\newblock {MetaGraph2Vec: Complex semantic path augmented heterogeneous network
  embedding}.
\newblock In {\em Lect. Notes Comput. Sci.}, pages 196--208. Springer, 2018.

\bibitem{graphSGD2018}
J.~X. Zheng, S.~Pawar, and D.~F. Goodman.
\newblock {Graph drawing by stochastic gradient descent}.
\newblock {\em IEEE Trans. Vis. Comput. Graph.}, 25(9):2738--2748, 2019.

\bibitem{Zhou2021}
Z.~Zhou, C.~Shi, X.~Shen, L.~Cai, H.~Wang, Y.~Liu, Y.~Zhao, and W.~Chen.
\newblock {Context-aware sampling of large networks via graph representation
  learning}.
\newblock {\em IEEE Trans. Vis. Comput. Graph.}, 27(2):1709--1719, 2021.

\bibitem{zhu2018deep}
D.~Zhu, D.~Wang, P.~Cui, and W.~Zhu.
\newblock {Deep variational network embedding in wasserstein space}.
\newblock In {\em KDD'18}. ACM, 2018.

\bibitem{Zhu2021}
M.~Zhu, W.~Chen, Y.~Hu, Y.~Hou, L.~Liu, and K.~Zhang.
\newblock {DRGraph: An Efficient Graph Layout Algorithm for Large-scale Graphs
  by Dimensionality Reduction}.
\newblock {\em IEEE Trans. Vis. Comput. Graph.}, 27(2):1666--1676, 2021.

\end{thebibliography}
\newpage

\end{document}
